\documentclass[prx,final,twocolumn,superscriptaddress,aps]{revtex4-2}
\usepackage{amsmath}
\usepackage{amssymb}
\usepackage[utf8]{inputenc}
\usepackage[colorlinks=True,linkcolor=blue,citecolor=blue,urlcolor=blue]{hyperref}
\usepackage{xcolor}
\usepackage{graphicx}
\usepackage{physics}
\usepackage{times}

\begin{document}
\title{Alternating quantum-emitter chains: \\ Exceptional-point phase transition, edge state, and quantum walks}
\author{Jimin Li}
\affiliation{Institute of Physics, University of Bonn, Nussallee 12, 53115 Bonn, Germany}
\author{Zongping Gong}
\affiliation{Department of Applied Physics, University of Tokyo, 7-3-1 Hongo, Bunkyo-ku, Tokyo 113-8656, Japan}
\affiliation{Theoretical Quantum Physics Laboratory, Cluster for Pioneering Research, RIKEN, Wako-shi, Saitama 351-0198, Japan}

\begin{abstract}
    We study the long-range hopping limit of a one-dimensional array of $N$ equal-distanced quantum emitters in free space, where the hopping amplitude of emitter excitation is proportional to the inverse of the distance and equals the lattice dimension. For two species of emitters in an alternating arrangement, the single excitation sector exhibits non-Hermitian spectral singularities known as exceptional points. We unveil an unconventional phase transition, dubbed \emph{exceptional-point phase transition}, from the collective to individual spontaneous emission behaviors. At the transition point, the $N \times N$ Hamiltonian fragments into $N/2-1$ many two-dimensional non-diagonalizable blocks. The remaining diagonalizable 
    block contains a dissipation-induced edge state with algebraically localized profiles, and we provide numerical evidence for its existence in the infinite-array limit. We demonstrate that the edge state can be eliminated via a continuous deformation, consistent with the ill-definedness of bulk topological invariant. We also propose a spatially resolved character to quantify the incoherent flow and loss in the non-unitary quantum walks of single atomic excitations.  
\end{abstract}

\maketitle
\section{Introduction}
\begin{figure}[t]
    \centering
    \includegraphics[width=1\columnwidth]{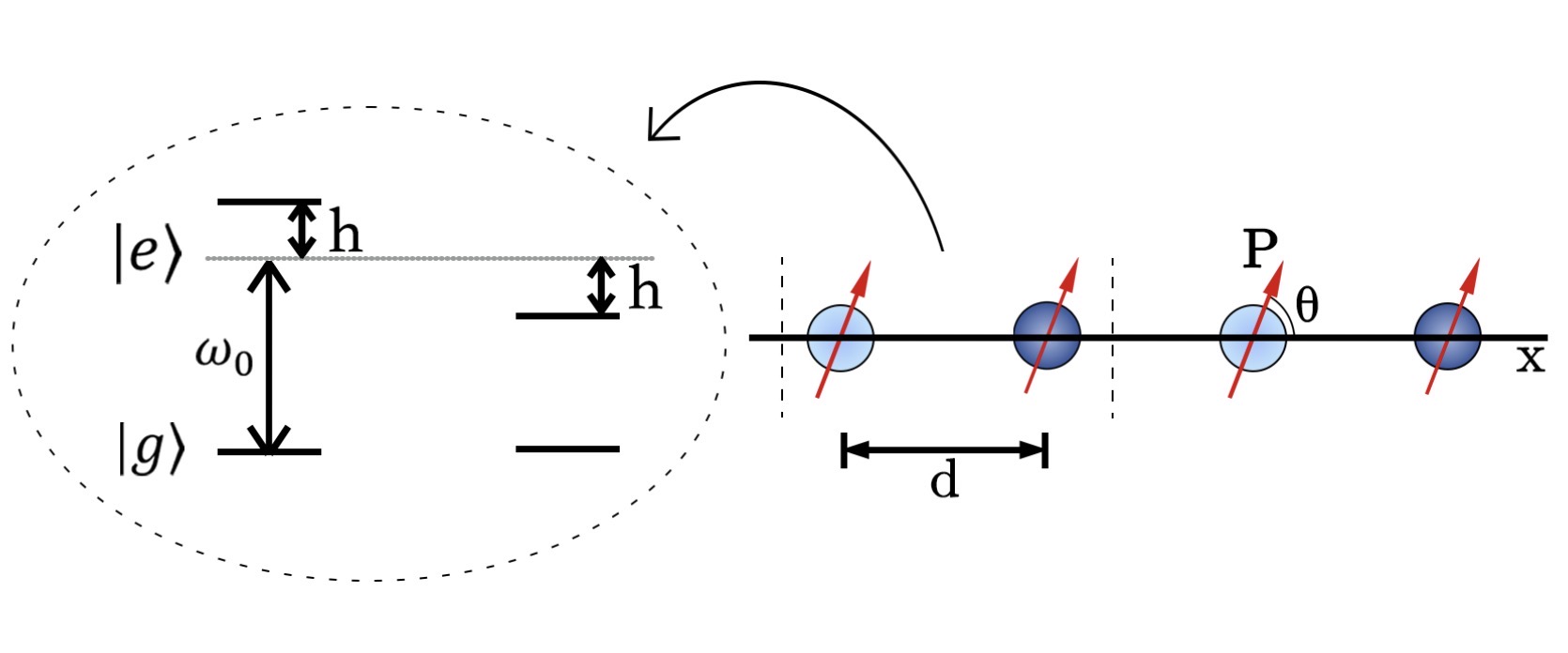}
    \caption{ Illustration of a 1D array of alternating quantum emitters in 3D free space, the two species have a transition energies $\omega_{0}\pm h$. The standard resonant dipole-dipole interaction (RDDI) is recovered by setting identical transition energies, namely $h=0$. 
     }  
    \label{fig:illustration}
\end{figure}
Structured arrays of quantum emitters, typically modeled as two-level atoms, have attracted much attention in the contexts of quantum technologies, photon storage, and non-linear optics, owning the advantages of having controllable light-matter interfaces~\cite{shahmoon_cooperative_2017,chang_quantum_2018,manzoni_optimization_2018,patti_controlling_2021, moreno_quantum_2021, srakaew_a_2023}. A main issue in improving the quantum protocols' efficiency is the phenomenon of spontaneous emission. In general cases, light couples to matter both coherently and dissipatively, causing photon leakage or re-absorbing into undesired optical modes. Recently, sub-radiant states of $N$ atoms with $\mathcal{O}(N^{-\alpha})$ decay rates were found in finite sub-wavelength atomic arrays in three-dimensional (3D) free space, known as the sub-radiance scaling~\cite{guerin_subradiance_2016, ferioli_storage_2021,sheremet_waveguide_2023,asenjo-garcia_exponential_2017, zhang_theory_2019}. Here sub-wavelength means that the emitter spacing is smaller than the light wavelength corresponding to the atomic transition frequency. Under such conditions, the light-mediated interactions between emitters can be described by an effective non-Hermitian long-range Hamiltonian, known as the resonant dipole-dipole interaction (RDDI)~\cite{asenjo-garcia_exponential_2017}.

Meanwhile, over the past years, the theoretical interests originated novel non-Hermitian phenomena such as the exceptional points (EPs)~\cite{ashida_non-hermitian_2020,heiss_physics_2012,bergholtz_exceptional_2021}, skin effects~\cite{yao_edge_2018,Okuma2023}, and non-Hermitian topology~\cite{gong_topological_2018,kawabata_symmetry_2019,Zhou2019} in various open classical and quantum systems that have synchronized with controllable experimental setups in the platforms of atomic, molecular and optical physics~\cite{gong_anomalous_2022,gong_bound_2022,roccati_hermitian_2023,xiao_observation_2020,xiao_observation_2023}. These phenomena arise from the unique complex spectral structures and non-orthogonality (or even incompleteness) of eigenstates without Hermitian counterparts. However, previous works mostly focus on short-range systems with finite hopping ranges. The interplay between non-Hermiticity and long-range interactions, which are the two fundamental features of sub-wavelength atomic arrays, remains largely unexplored.

In this work, we study the single excitation sector of sub-wavelength alternating atomic chains in 3D %three-dimensional (3D) 
free space through the lens of non-Hermitian physics. In particular, we consider the hopping decay power equal to the lattice dimension, which is a marginal value from the angle of thermodynamic stability~\cite{Kuwahara2021,gong_long-range_2023}. When tuning the strength of alternation, we demonstrate this realistic model shows a novel non-Hermitian behavior which we call \emph{exceptional-point phase transition}. At the transition point, the $N\times N$ Hamiltonian in the single excitation sector can be decomposed into $N/2-1$ many non-diagonalizable (Jordan) and $1$ diagonalizable $2 \times 2$ blocks. Away from the transition point, the new dispersion relation modifies the RDDI sub-radiance scaling.

In addition, we numerically show the existence of an edge state with a power-law decaying tail under the dissipative long-range hopping, despite the fact that a bulk topological invariant is ill-defined. Lastly, we consider the real-time dynamics by investigating a spatially resolved escape distribution and wavefunction density. We generalize the conventional character to account for not only on-site loss but also incoherent hopping. We find an initial quantum walker with parity symmetry shows an imbalance of escape rates and densities at the boundaries at late time. Our work provides a paradigm for exploring novel phases and dynamics in long-range non-Hermitian systems, for which na\"ive approximations such as dropping of non-Hermitian terms and truncation of long-range hopping may lead to qualitatively incorrect results.

\section{1D array of Quantum Emitters}
We consider a 1D array of $N$ equal distanced quantum emitters in 3D free space. Each emitter is a two-level atom with an internal structure of the ground $\ket{g}$ and excited state $\ket{e}$, separated by an atomic transition energy gap $\omega_{0}$ (see in Fig. \ref{fig:illustration}). For a sub-wavelength interatomic distance $d$, the collective spontaneous emission of atomic ensemble involves light-matter interactions between atoms and all the electromagnetic modes in free space. Welsch et al. have developed a generalized input-output formalism based on the classical 3D free-space electromagnetic Green's tensor $\mathbf{G}\left(\mathbf{r}_{i},\mathbf{r}_{j},\omega\right)$ to capture such collective effect, where $\mathbf{r}_i$ denotes the position of the $i$th atom~\cite{lehmberg_radiation_1970, gruner_green-function_1996, xu_input-output_2015, shi_multi-photon_2015}. Under the Born-Markov approximation $\mathbf{G}\left(\mathbf{r}_{i},\mathbf{r}_{j},\omega\right) \simeq \mathbf{G}\left(\mathbf{r}_{i},\mathbf{r}_{j},\omega_{0}\right)$, one can solve for the atomic degrees of freedom and obtain an effective light-field mediated long-range non-Hermitian Hamiltonian, i.e., the RDDI Hamiltonian
\begin{equation}
H_{\text{RDDI}} = - \mu_{0} \omega_{0}^{2} \sum_{ i\neq j; i,j=1}^{N}   \mathbf{P^{*}}  \cdot \mathbf{G}\left(\mathbf{r}_{i},\mathbf{r}_{j},\omega_{0}\right)  \cdot \mathbf{P}\: \sigma^{\dagger}_{i}\sigma_{j},
\label{eq:H_RDDI}
\end{equation}
where $\mu_0$ is the magnetic constant, $\sigma^{\dagger}_{i} = |e_{i} \rangle \langle g_{i}|$ excites the $i$th atom, $\mathbf{P}$ is the transition dipole, and $\mathbf{G}\left(\mathbf{r}_{i},\mathbf{r}_{j},\omega_0\right)=\mathbf{G}\left(\mathbf{r}_{i}-\mathbf{r}_{j},\omega_0\right)$ with
\begin{equation}
    \begin{split}
\mathbf{G}\left(\mathbf{r},\omega_{0}\right) = &\frac{e^{i k_{0}r}}{4 \pi k_{0}^{2}r^{3}} \Bigl[ \left( k_{0}^{2}r^{2} + i k_{0}r - 1 \right) \mathbf{1} \\ 
&+ \left( -k_{0}^{2}r^{2} - 3ik_{0}r + 3 \right) \frac{\mathbf{r}\otimes\mathbf{r}}{r^{2}} \Bigr],
\end{split}
\end{equation}
$k_{0} = \omega_{0}/c$, $r = |\mathbf{r}|$ ($c$: speed of light). %\abs{\mathbf{r}_{i}-\mathbf{r}_{j}} 
 Hereafter, we %further 
 set the atomic spontaneous emission rate $\gamma_0=\mu_0\omega_0^3|\mathbf{P}|^2/(3\pi\hbar c)$ to be the unit. 
 
In the $N\to\infty$ infinite-array limit and under the assumption that the coordinate origin coincides with an atom, the single excitation sector of $H_{\text{RDDI}}$ is diagonalized by the Bloch states
\begin{equation*}
H_{\text{RDDI}} \left( \sigma^{\dagger}_{\mathbf{k}}\ket{\mathbf{g}} \right) =\omega_{\text{eff}} \left( \sigma^{\dagger}_{\mathbf{k}}\ket{\mathbf{g}} \right),
\end{equation*}
where $\ket{\mathbf{g}}=\ket{g}^{\otimes N}$, $\sigma^{\dagger}_{\mathbf{k}} = \sum^{N}_{j=1}e^{i\mathbf{k} \cdot \mathbf{r_{j}}}\sigma^{\dagger}_{j}/\sqrt{N}$, and
\begin{equation*}
\omega_{\text{eff}} = -\frac{3\pi}{k_{0}}\hat{\mathbf{P}}^{*} \cdot \left[ \sum^{N}_{j=1}e^{-i\mathbf{k} \cdot \mathbf{r_{j}}} \mathbf{G}\left(\mathbf{r}_{j}, \omega_{0}\right) \right] \cdot \hat{\mathbf{P}}
\end{equation*}
is the dispersion relation (here $\hat{\mathbf{P}}=\mathbf{P}/|\mathbf{P}|$), which is determined from the Fourier-transformed Green's tensor~\cite{asenjo-garcia_exponential_2017}. The long-range hopping of the excitation forms a band that has the energy $\Re(\omega_{\text{eff}})$ and decay rate $-2\Im(\omega_{\text{eff}})$. Recalling that all the atoms are located on a line, which is chosen to be the $x$ axis, we know that the perpendicular component of $\mathbf{k}$ does not alter the state or energy. Hence, it suffices to focus on $\mathbf{k}=k\hat{\mathbf{x}}$ with $k\in[-\pi/d,\pi/d]$. The Bloch states of $|k|>k_{0}$ are off-resonant and hence perfectly sub-radiant, namely $\Im(\omega_{\text{eff}})=0$. 

Realistic experimental realizations of such a 1D array consist of only a finite number of atoms with an open boundary condition (OBC), which modifies the infinite chain results. The major effect is that all the off-resonant states gain a non-zero decay rate; Starting from the lowest decay rate, the first $\xi / N \ll 1$ of sub-radiant states show a $N^{-\alpha}$ scaling and recover a zero decay rate in the infinite-array limit. Recently, the power law exponent $\alpha$ for the sub-radiant states was unraveled being the power-law scaling behavior of group velocity $\partial \omega_{\text{eff}}/\partial k$ near the band edge $k = \pm \pi/d$~\cite{zhang_subradiant_2020-1} and closely related to the overlap error between a set of finite chain Ansatz and the Bloch states~\cite{asenjo-garcia_exponential_2017}. However, such scaling and Ansatz break down near the light line $k=\pm k_0$~\cite{zhang_theory_2019,zhang_subradiant_2020-1}.

\subsection{Long-range hopping limit}\begin{figure}[t]
    \centering
    \includegraphics[width=1\columnwidth]{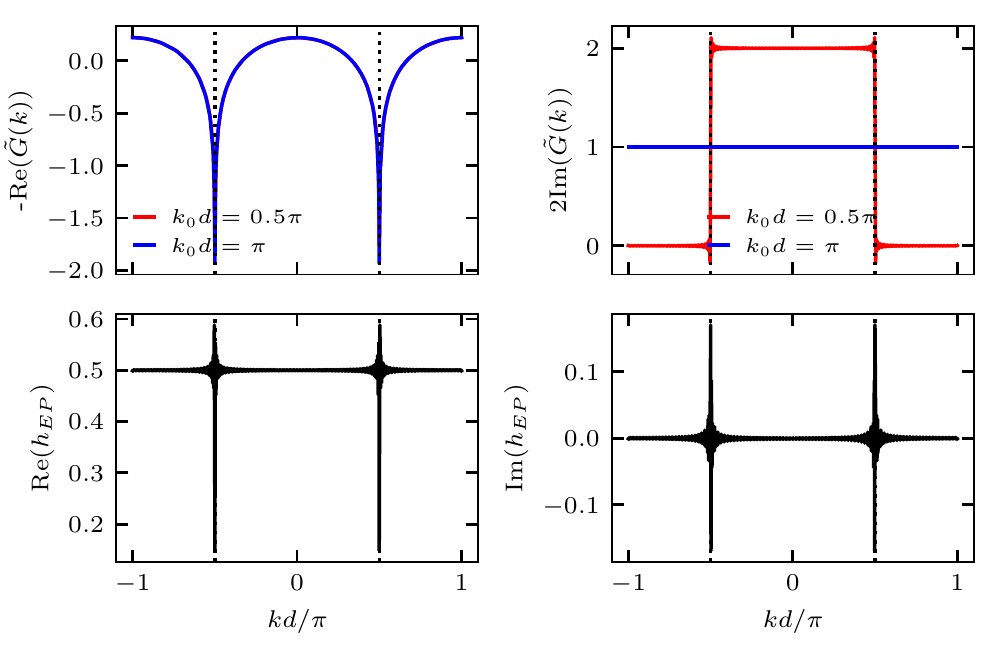}
    \caption{ (Upper) The discrete Fourier transformed Green's function (\ref{eq:Gdk}) 
    in 1D at $ dk_{0} = \pi/2$(red) and $ dk_{0} = \pi$(blue) for the long-range hopping $\theta=\arccos(1/\sqrt{3})$. The imaginary part of the red curve is discontinuous at the light line. (Lower) The infinite-array limit predictions of $h_{EP}$ at $\theta%_{\text{EP}}
    =\arccos(1/\sqrt{3})$. $h_{EP}$ equals 0.5 for states of all quasi-momentum except the light line.}
    \label{fig:twoband_h_EP}
\end{figure} 
In 1D atomic arrays, the hopping terms 
\begin{equation}
G(r)=\mu_0\omega_0^2\mathbf{P}^*\cdot \mathbf{G}( r\hat{\mathbf{x}},\omega_0)\cdot\mathbf{P}
\label{eq:Gr}
\end{equation}
of $H_{\text{RDDI}}$ are in the following power-law forms \cite{reitz_cooperative_2022}: 
\begin{multline*}
 \Re \bigl( G(r) \bigr)
  \propto \left( 1 - \cos^2\theta \right) \frac{\cos(k_{0}r)}{k_{0}r} \\ - \left( 1 - 3\cos^2\theta \right)\left[ \frac{\sin(k_{0}r)}{(k_{0}r)^{2}} + \frac{\cos(k_{0}r)}{(k_{0}r)^{3}}\right],
\end{multline*}
\begin{multline*}
\Im \bigl( G(r) \bigr)
 \propto \left( 1 - \cos^2\theta \right) \frac{\sin(k_{0}r)}{k_{0}r} \\ - \left( 1 - 3\cos^2\theta \right)\left[ \frac{\cos(k_{0}r)}{(k_{0}r)^{2}} - \frac{\sin(k_{0}r)}{(k_{0}r)^{3}}\right], 
\end{multline*}
where the highest power exponent can be $r^{-1}$ or higher depending on the angle from dipole to the 1D array $\theta=\arccos ( \hat{\mathbf{P}}\cdot\hat{\mathbf{x}} )$.
%(\mathbf{P}\cdot\mathbf{e}_x/|\mathbf{P}|)
In this work, we focus on the long-range hopping limit of $H_{\text{RDDI}}$ by setting $\theta = \arccos(1/\sqrt{3}) =  0.955$, where the power-law decay hopping exponent equals the spatial dimension and the resulting term are simply proportional to the zeroth spherical Bessel functions.

Upper panels in Fig. \ref{fig:twoband_h_EP} (red) show the dispersion relation in the infinity chain limit at an interatomic distance $dk_{0}=\pi/2$, obtained by the discrete Fourier transform (FT) of the Green's function (\ref{eq:Gr}). The decay rate is given by the rectangle function, which is just the FT of $\sin(r)/r$. Thus, the collective decay rate is a constant for all $|k|<\pi/(2d)$ within the light line, discontinuous at the light line $|k|=\pi/(2d)$ and zero for $|k|>\pi/(2d)$, implying perfect sub-radiance. The real part of $H_{\text{RDDI}}$ is proportional to $\cos(k_{0}r) / (k_{0}r)$, and its FT is ill-defined due to the singularity at $r=0$. Such diverging on-site potential is forbidden here, and the discrete FT 
\begin{equation}
\tilde G_d(k) = \sum_{n\in\mathbb{Z}\backslash\{0\}} e^{-ikdn}G(dn)
\label{eq:Gdk}
\end{equation}
has no closed form. 

For reasons to be clear soon, we also consider the discrete Fourier transformed Green's function $\tilde{G}_{2d}(k)$ at another interatomic distance $dk_0=\pi$, which is the minimal interatomic distance for the absence of sub-radiance in 1D finite array and the dispersion is shown in Fig. \ref{fig:twoband_h_EP} (blue). We find the numerical values of the real dispersion are the same as $\Re(\tilde{G}_{d}(k))$ up to negligible errors. The imaginary part of the dispersion is $1$ for all $k$, reflecting that all emitters search the atomic limit with the decay rate equal to the spontaneous emission without a light line, i.e. collective emission is absent. 

\subsection{Alternating Quantum Emitters }
Having introduced the properties of long-range hopping sub-wavelength atomic arrays, we consider a two-band model -- a 1D array of two different alternating arranged species of quantum emitters with transition energies $\omega_{0} \pm h$ in free space, as illustrated in Fig. \ref{fig:illustration}~\cite{reitz_cooperative_2022}. The corresponding %RDDI 
Hamiltonian 
\begin{equation}
H_{\text{TB}} = H_{\text{RDDI}} + h\sum_{j=1}^{N}(-1)^{j+1}\sigma^{\dagger}_{j}\sigma_{j}
\label{eq:twoband}
\end{equation}
is block-diagonalizable in the quasi-momentum space in the infinite-array limit. To see this, we write the long-range hopping terms explicitly 
\begin{equation*}
H_{\text{TB}} = -\sum^\infty_{i=-\infty}\sum_{n\in\mathbb{Z}\backslash\{0\}}
G(nd)\sigma^{\dagger}_{i+n}\sigma_{i} +  h\sum_{j=-\infty}^{\infty}(-1)^{j+1}\sigma^{\dagger}_{j}\sigma_{j}
\end{equation*}
and introduce the Fourier transformed spin operators $o^{\dagger}_{k} = \sqrt{2/N}\sum_{j=1}^{N/2}e^{ikr_{2j-1}} \sigma^{\dagger}_{2j-1}$, $e^{\dagger}_{k} = \sqrt{2/N}\sum_{j=1}^{N/2}e^{ikr_{2j}} \sigma^{\dagger}_{2j}$ and $dk \in [-\pi/2,\pi/2]$ for the odd and even sites. Rewriting Eq. (\ref{eq:twoband}) in the quasi-momentum space leads to
\begin{equation}
H_{\text{TB}} = \sum_{k} 
\begin{pmatrix}
o^{\dagger}_{k} & e^{\dagger}_{k} \\
\end{pmatrix}
\tilde{H}_{\text{TB}}(k)
\begin{pmatrix}
o_{k} \\ e_{k}
\end{pmatrix},
\end{equation}
where
\onecolumngrid
\begin{equation*}
\tilde{H}_{\text{TB}}(k)=
\begin{pmatrix}
h - \sum_{n\in\mathbb{Z}\backslash\{0\}}%\sum_{n \in \mathbb{Z}_{\neq 0}}^{\pm\infty} 
e^{-ik2dn}G(2dn) & -e^{-ikd}\sum^\infty_{n=-\infty}e^{-i(2n-1)kd}G
\Bigl( (2n-1)d \Bigr)%\sum^{\pm\infty}_{n \in \pm1,\pm3, \dots} e^{-ikdn} G \left(dn\right) 
\\
-e^{ikd}\sum^\infty_{n=-\infty}e^{-i(2n-1)kd}G\Bigl((2n-1)d\Bigr) %\sum^{\pm\infty}_{n \in \pm1,\pm3, \dots}  e^{-ikdn} G\left(dn\right) 
& -h - \sum_{n\in\mathbb{Z}\backslash\{0\}}%\sum_{n \in \mathbb{Z}_{\neq 0}}^{\pm\infty} 
e^{-ik2dn}G(2dn)\\
\end{pmatrix}.
%\begin{pmatrix}
%o_{k} \\ e_{k}
%\end{pmatrix}.
\end{equation*}
Noting that the sum in the off-diagonal component is  
\begin{equation*}
%e^{-ikd}
%\sum^{\pm\infty}_{n \in \pm1,\pm3, \dots}e^{-ikn} G\left(dn\right) 
\sum^\infty_{n=-\infty}e^{-i(2n-1)kd}G\Bigl((2n-1)d \Bigr)= %e^{-ikd} \left(  
\sum_{n\in\mathbb{Z}\backslash\{0\}}%\sum^{\pm\infty}_{n \in \mathbb{Z}_{\neq 0}} 
\Bigl( e^{-ikdn} G\left(dn\right) -  e^{-ik2dn} G\left(2dn\right)\Bigr) %\right) 
= %e^{-ikd} \left( 
\tilde{G}_{d}(k) -  \tilde{G}_{2d}(k)  %\right)
, 
\end{equation*}
we obtain the following simple form for the two-band Bloch Hamiltonian:
\begin{equation}
\label{eq:twoband_kspace}
\tilde{H}_{\text{TB}}(k) = \begin{pmatrix}
h  - \tilde{G}_{2d}(k) & -e^{-ikd} \left( \tilde{G}_{d}(k) -  \tilde{G}_{2d}(k)  \right) \\
-e^{ikd} \left( \tilde{G}_{d}(k) - \tilde{G}_{2d}(k) \right) & -h - \tilde{G}_{2d}(k) 
\end{pmatrix}.
\end{equation}
\twocolumngrid

\section{Exceptional-point phase transition}
\begin{figure}[t]
    \centering
    \includegraphics[width=1\columnwidth]
    {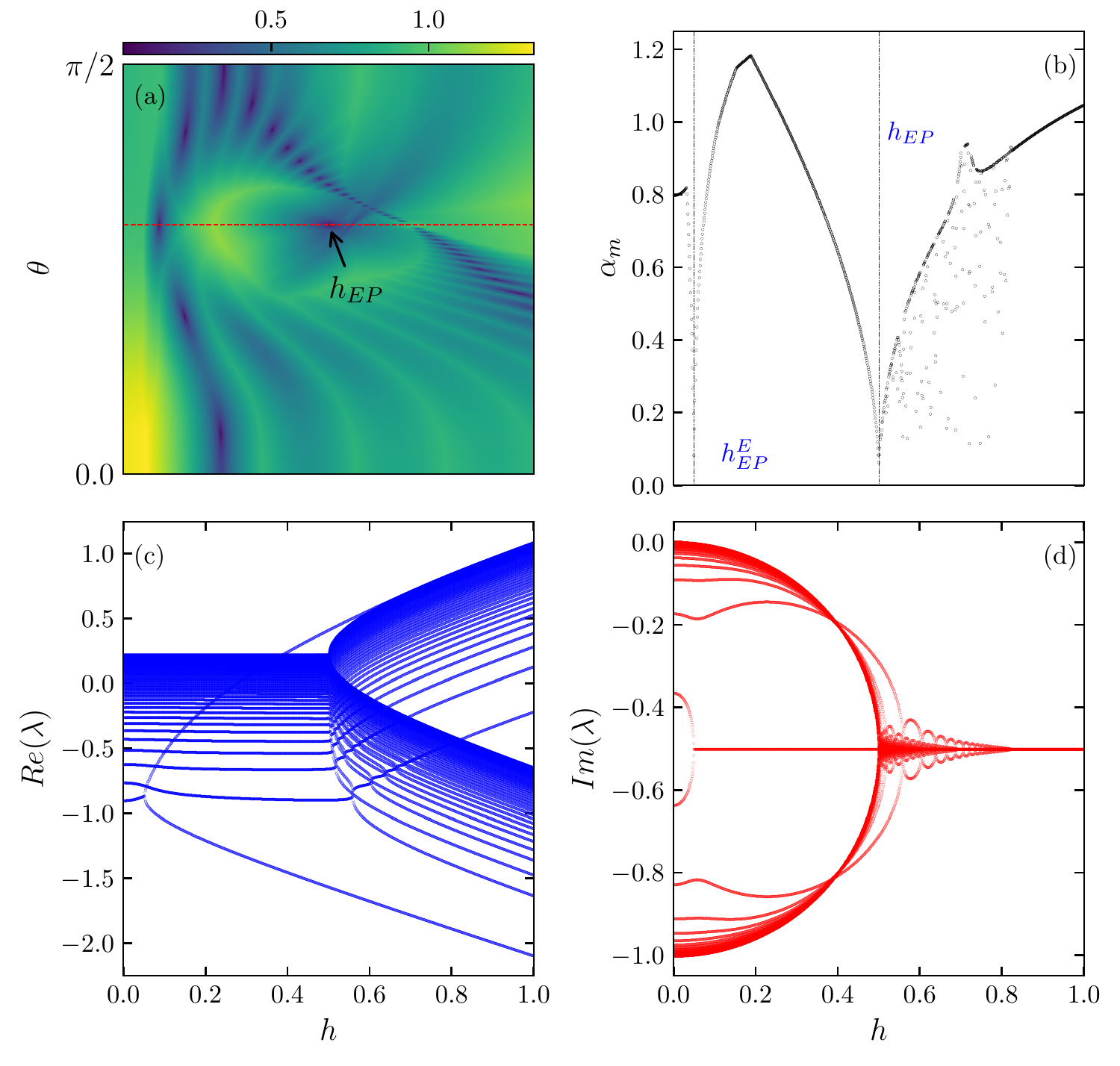}
    \caption{ Finite chain OBC calculation of (a) the smallest angle between all pairs of right eigenvectors for $N=40$ and all possible $\theta, h$. The red dotted line indicates the long-range hopping $\theta$ considered in the main text. EPs are commonly found in the parameter space~\cite{luitz_exceptional_2019}. (b) The smallest angle between all pairs of right eigenvectors %Eq. 
    (\ref{eq:metric}) for $N=100$. It signals the predicted EP phase transition at $h=0.5$ and finite size induced $h^{E}_{EP}$ of the edge state Jordan block. (c,d) The real and imaginary parts of the spectrum for $N=100$. For $h<0.5$, the spectrum is imaginary line-gapped. In the infinite-array limit, the system contains $N-1$ two-level EPs at the EP transition point $h=0.5$. The finite chain exact diagonalization spectrum gives a good agreement for states away from the light line.}
    
    \label{fig:twoband_heatmap}
\end{figure}

The spatially alternating transition energies divides $N$ atoms into two sublattices and leads to the formation of two bands with a dispersion relation
\begin{equation}
\omega^{\pm}_{\text{eff}}(k) = -\tilde{G}_{2d}(k) \pm \sqrt{ \left( \tilde{G}_{d}(k) - \tilde{G}_{2d}(k) \right)^{2}  + h^{2} }
\label{eq:omega}
\end{equation}
in the infinite-array limit. A necessary condition for the emergence of EPs is 
\begin{equation}
-h^{2} = \left( \tilde{G}_{d}(k) - \tilde{G}_{2d}(k) \right)^{2},
\label{eq:EP_condition}
\end{equation}
which relates the differences in transition energies to the dispersions in the identical array limit $h=0$. Recall that the collective decay rate is a constant in the long-range hopping limit (cf. Fig.~\ref{fig:twoband_h_EP}), and the light line 
coincides with the new first Brillouin zone edge $k = \pm \pi / (2d)$. The real parts of the dispersion at both interatomic distances experience a cusp at $k = \pm \pi/(2d)$ and match exactly for other $k$. Therefore, the two-band Hamiltonian is non-diagonalizable for every $k \neq \pm \pi /(2d)$ at $h_{EP} = 0.5$ from Eq.~(\ref{eq:EP_condition}). Due to this peculiar feature that the \emph{whole} system (except the band edges) undergoes an EP transition, we call the phenomenon \emph{EP phase transition}. Physically speaking, as $h$ increases, the spontaneous emission mechanism changes from collective to individual in a square-root fashion (see Fig.~\ref{fig:twoband_heatmap}(d)). 

The EP phase transition (collective-atomic emission transition) is analogous to the metal-insulator transition in the Aubry-Andr\'e-Harper (AAH) model \cite{Harper1955,Aubry1980}, where the localization transition occurs \emph{simultaneously} at the full energy spectrum. In the usual short-range $\mathcal{PT}$-symmetric 1D models \cite{tzortzakakis_topological_2022, weimann_topologically_2017}, the EP is analogous to the mobility edge in a generalized AAH model~\cite{biddle_predicted_2010, ganeshan_nearest_2015,Liu2020}, and the (de)localized regions correspond to $\mathcal{PT}$-broken(symmetric) regions. 

Figure \ref{fig:twoband_heatmap} shows the finite chain spectrum for $N=100$ to confirm the infinite chain prediction, there are $N-1$ EPs near $h=h_{EP}=0.5$. Furthermore, it is useful to consider the smallest angle between all pairs of normalized right eigenvectors
\begin{equation}
\alpha_{m}= \min_{i\neq j} \arccos\left( | \langle \psi^{R}_{i}| \psi^{R}_{j}\rangle| \right),
\label{eq:metric}
\end{equation}
which goes to zero toward EPs and remains $\pi/2$ for diagonalizable spectrum degeneracies~\cite{schafer_symmetry_2022}. 

The angle metric also detects an EP at $h^{E}_{EP} \approx 0.05$ due to the finite size effect. This $2\times2$ block has the lowest real eigenvalues that match the $k=\pm \pi/(2d)$ states in Fig. \ref{fig:twoband_h_EP}, which we identify with being the span. Note that the infinite-array limit of $\tilde{H}_{\text{TB}}(k=\pm \pi/(2d))$ is always diagonal, because $\tilde G_d(k)$ is discontinuous and coincides with $\tilde G_{2d}(k)$ at the light line. The open chain correction consists of a Fourier sum error of $\sin(r)/r$ that is bounded by $1/N$ known as the Gibbs phenomenon~\cite{riley_mathematical_2006}, and systematic finite size error when approximating the true finite array eigenstates by the Bloch states. The finite-size Ansatz error scaling of sub-radiance is discussed in Refs.~\cite{asenjo-garcia_exponential_2017, zhang_theory_2019, zhang_subradiant_2020-1}, but such behavior is absent for states in the middle of the decay spectrum, which stops us from gaining an analytic finite size scaling. We numerically find that $h^{E}_{EP}$ goes to zero in the infinite chain limit with a $N^{-0.65}$ power-law scaling. 

In addition, we find the position of EP around $h \approx 0.55$ remains for larger $N$. The quasi-momentum of this state in the $h=0$ limit is the nearest $k$ from the light line, and it is delocalized in the real space. Due to the presence of discontinuity, the finite chain results deviate from the EP phase transition picture for states near the light line. 

The new dispersion relation Eq. (\ref{eq:omega}) implies a change in sub-radiance scaling for small differences in transition energies. For $h=0$, such scaling of Eq. (\ref{eq:H_RDDI}) is $N^{-3}$ due to the quadratic dispersion at the band edge, as the smallest non-vanishing expansion order in $k$ of $\Re(\tilde G_d(k))$ near $k=\pm\pi/d$ is 2 \cite{zhang_subradiant_2020-1}. Similarly, the dispersion of the two-band model is given by Eq. (\ref{eq:omega}), and expansion in small $h$ yields a new overall sub-radiance scaling, $\mathcal{O}(N^{-3})+\mathcal{O}(h^2)$.

\begin{figure}[t]
    \centering
    \includegraphics[width=1\columnwidth]{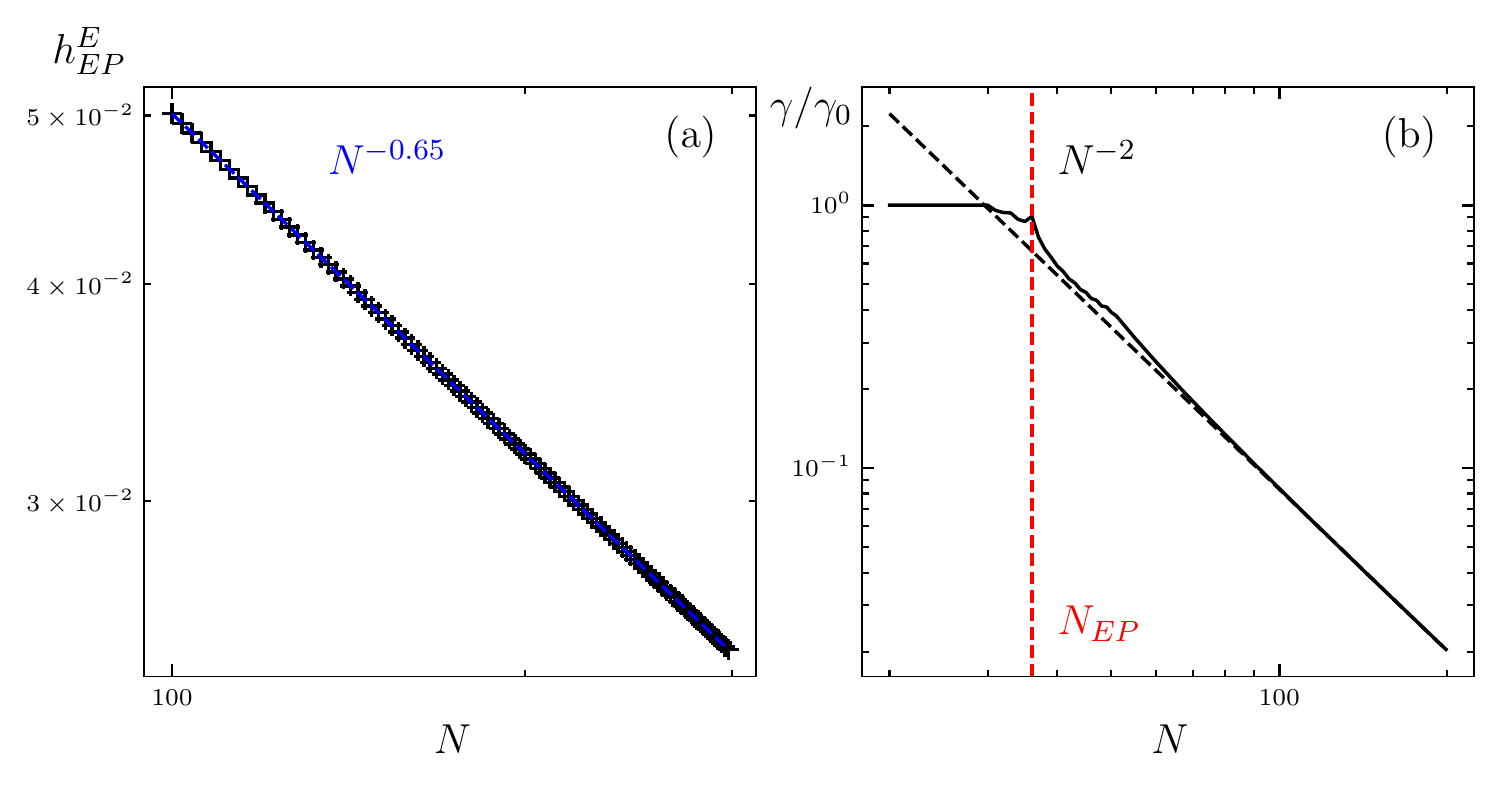}
    \caption{ Exact diagonalization results for (a) the finite size scaling of $h^{E}_{EP}$. The system shows a localized edge state for arbitrary small $h$ greater than $N^{-0.65}$. (b) The sub-radiance scaling for $h = 20/N$. The ED result is captured by the infinite chain dispersion given in Eq. (\ref{eq:omega}).
    This implies that small imperfection in transition energies would change the RDDI $N^{-3}$ sub-radiance scaling.}
    \label{fig:paper_subradiance_scaling}
\end{figure} 
Thus, we take $h = N^{-\alpha}$ to observe a valid scaling, and collective emission is suppressed for any $h> h_{EP}=0.5$. Evidently, the RDDI scaling is recovered if $\alpha \ge 3/2$. Otherwise, %And 
the scaling is dominated by $h$ if $\alpha < 3/2$. Figure \ref{fig:paper_subradiance_scaling} shows the scaling transition for $h = 20/N$. The most sub-radiant decay rate is in the same order as the spontaneous emission rate $\gamma_{0}$ for $N<N_{EP}=40$, i.e., $h\geq0.5$ when $N \leq 40$. Further system size increases show the predicted $N^{-2}$ scaling. Note that previous studies found that small disorder in atomic arrays suppresses the usual RDDI sub-radiant ($N^{-3}$) scaling~\cite{kornovan_extremely_2019}. Here, in contrast, we find a small alternative on-site potential makes the system more radiative.

\section{Edge State}
\begin{figure}[t]
    \centering
    \includegraphics[width=1\columnwidth]{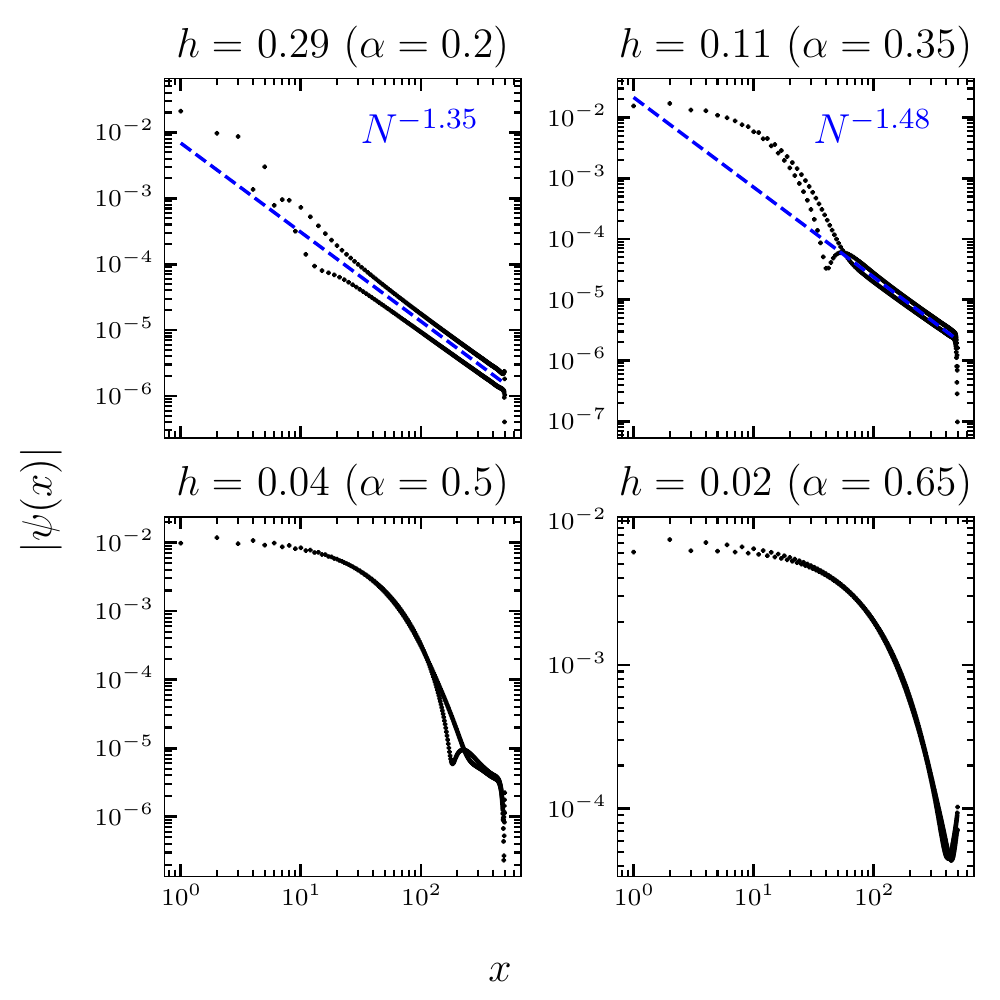}
    \caption{Real-space distribution $|\psi(x)|$ of edge state at 4 different $h=N^{-\alpha}$ for $N=500$. For $h<h^{E}_{EP}$, the localization length of the edge state is extensive in system size. Once $h$ overcomes the finite size threshold $h^E_{EP}$, the edge state becomes localized and shows a clear power-law decaying tail owing to the long-range nature of the model.}
    \label{fig:edge}
\end{figure}

Having discussed the coalescence of the $k=\pm \pi/(2d)$ states at $h^{E}_{EP}$, we consider their real-space localization properties, as shown in  Fig. \ref{fig:edge} for a range of $h = N^{-\alpha}$. In a finite chain of $N=500$ emitters, the spatial distribution $|\psi(x)|$ is localized at one end with the atomic spontaneous emission rate. We identify this mode as a dissipation-induced edge state, in the sense that such localization is not observed when discarding the dissipative part of the Hamiltonian. %\zpg{The number of edge states is one(two) for $N$ being even(odd) [ZG: Maybe not necessary to mention this since we already specify $N=500$ (another reason is that a unit cell contains two atoms so an appropriate OBC implies the total atom number to be even); otherwise we should explain why this difference appears]}. 
For $h < h^{E}_{EP}$ (Fig. \ref{fig:edge} Lower), we find the localization length of this wavefunction to be extensive in system size, such that its amplitude decreases to zero roughly at the opposite end of the chain. For $h>h^{E}_{EP}$, the real-space amplitude is truly localized and shows an algebraic decay tail, which has been numerically observed in other long-range models~\cite{Vodola_long-range_2016, jager_edge_2020}.  

It is tempting to attribute the edge state to the nontrivial bulk properties of the model. Indeed, the number of edge states of a Hermitian short-range Hamiltonian is related to its bulk topological number, 
known as the bulk-edge correspondence, which has been generalized to non-Hermitian systems~\cite{lee_anomalous_2016,ashida_non-hermitian_2020,yao_edge_2018,kunst_biorthogonal_2018,Slager2020}. However, the topological properties of $\tilde{H}_{\text{TB}}(k)$ is ill-defined due to the discontinuity of $\tilde{G}_{d}(k)$ in $k$ %Eq.~(\ref{eq:twoband_kspace}) 
as a manifestation of the long-range hopping~\cite{gong_long-range_2023}. In the following, we explain the ill-definedness in further detail and show that the edge state is \emph{not} topologically protected, as it can be removed via a symmetry-preserving continuous deformation.  

We consider the two-band model in the Su–Schrieffer–Heeger-like (SSH-like) form by rewriting Eq. (\ref{eq:twoband_kspace}) in terms of the spin-$\frac{1}{2}$ Pauli matrices $\sigma_{0,x,y,z}$ with $\tilde{G}_{d}(k) -  \tilde{G}_{2d}(k) = ig(k)$, where 
\[
  g(k) =
  \begin{cases}
    h_{EP} & \text{$k < | \frac{\pi}{2d} |$} \\
    0 & \text{ $k = \pm \frac{\pi}{2d}$} \\
    -h_{EP} & \text{$k > | \frac{\pi}{2d} |$}
  \end{cases}
\]
is real. Note that $g(k)$ is a \emph{discontinuous} function and changes its sign upon crossing the light line. This sign change ensures the periodicity $\tilde H_{\rm TB}(k+\pi/d)=\tilde H_{\rm TB}(k)$. Within $kd\in[-\pi/2,\pi/2]$ ,
$\tilde{H}_{\text{TB}}(k)$ reads %\zpg{We can replace $g(k)$ by $h_{EP}$ since $k\in(-\pi/(2d),\pi/(2d))$}
\begin{equation*}
\begin{split}
\tilde{H}_{\text{TB}}(k) &= -\tilde{G}_{2d}(k)\sigma_0 + h \sigma_{z} \\
&- ig(k)\cos(kd)\sigma_{x } - ig(k)\sin(kd)\sigma_{y}. 
\end{split}
\end{equation*} 
Strictly speaking, the spectrum is \emph{separable} for a non-vanishing $h$~\cite{shen_topological_2018}. Without the discontinuity, the spectrum is imaginary \emph{line-gapped} at $\Im\omega= \Im \tilde{G_{d}}(k=\pm \pi/(2d))$ for $h<h_{EP}$, and separable for $h>h_{EP}$.

Without loss of generality, the $\sigma_0$ term can be dropped since $\tilde H_{TB}(k)$ can be continuously connected to a Hamiltonian $\tilde{H}_{\text{TB}}'(k)$ without $\sigma_0$:
\begin{equation}
\tilde{H}_{\text{TB}}'(k) = h \sigma_{z} - ig(k)\cos(kd)\sigma_{x } - ig(k)\sin(kd)\sigma_{y}. 
\label{eq:twoband_SSH}
\end{equation} 
This model appears to be similar to a short-range $\mathcal{PT}$-symmetric SSH model~\cite{tzortzakakis_topological_2022}. Despite the similarity, the latter has a pair of edge states at two ends of the chain, and they are topologically protected. As shown in Appendix A, Hermitization of the short-range SSH model leads to a well-defined winding number $\mathbb{Z}$. In the former case, the range of quasi-momentum is only defined as half of the Brillouin zone $dk \in [-\pi/2,\pi/2]$ -- a manifestation of long-range hopping. Therefore, the Bloch Hamiltonian is discontinuous and only one of the $k=\pm \pi/(2d)$ states is an edge state while another is delocalized. 
\begin{figure}[t]
    \centering
    \includegraphics[width=1\columnwidth]{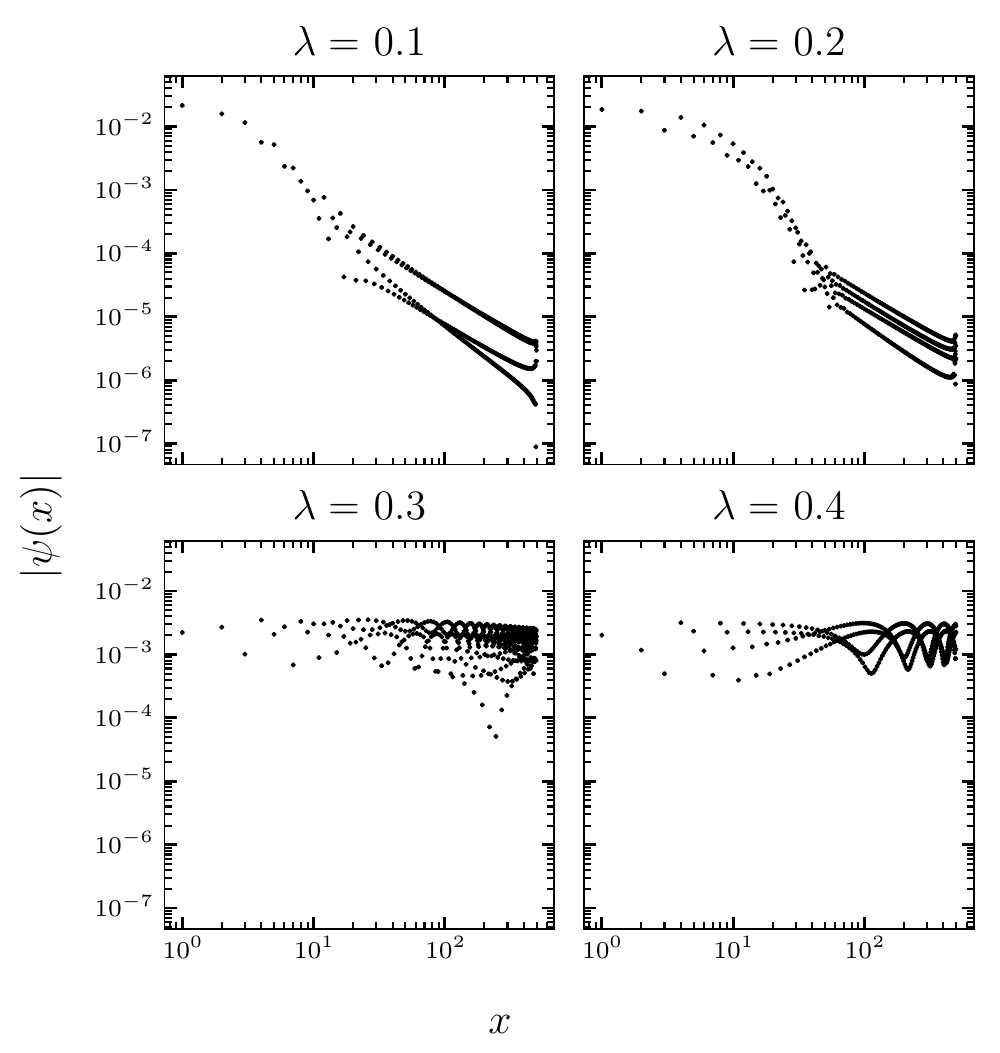}
    \caption{Edge state of the deformed Hamiltonian %Eq.
    (\ref{eq:deformation}). The real space amplitude $|\psi(x)|$ is shown at 4 different $\lambda$ for $N=500$ and $h=N^{-0.25} \approx 0.21$. Increasing $\lambda$ delocalizes the power-law tailed edge state, which indicates that it is not symmetry-protected.} 
    \label{fig:deformation}
\end{figure} 

Since the edge state does not correspond to a well-defined bulk property, we consider a continuous deformation of Eq. (\ref{eq:twoband_SSH}) to a topologically trivial Hamiltonian 
\begin{equation}
H(\lambda) = (1 - \lambda )\tilde{H}_{\text{TB}}'(k) + \lambda H'' \quad \lambda \in [0,1],
\label{eq:deformation}
\end{equation}
where $H'' = i \sigma_{x}$ such that both the pseudo-Hermiticity and imaginary line-gap are preserved along the path. Figure~\ref{fig:deformation} shows the real space amplitude of the edge state at four different $\lambda$, where increasing $\lambda$ gradually delocalizes the edge state. 

\section{Quantum Walks}In the last section, we consider the real-time evolution of an initial state $\ket{\psi_{0}}$ under $H_{\text{TB}}$, known as the continuous-time quantum walk~\cite{Childs2010}. Under such non-unitary time evolution, the spontaneous emission occurs during $[t,t+dt]$ with a probability $ -2\bra{\psi(t)}\Im(H_{\text{TB}})\ket{\psi(t)}dt$, where $\ket{\psi(t)} = e^{-iH_{\text{TB}}t}\ket{\psi_0}$. We are interested in how the spontaneous emission can be resolved spatially.

Owning the experimental developments of spatially local quantum controls \cite{haroche_6exploring_2006}, previous works have mostly focused on non-Hermitian Hamiltonians with on-site loss \cite{xue_non-hermitian_2022}. The anti-Hermitian part of such a Hamiltonian is diagonal in the real space. Hence, the emission (escape) probability from a site $x$ is proportional to $\int^{\infty}_{0} |\langle x | \psi(t) \rangle |^{2}dt$, where $| x \rangle = \sigma^{\dagger}_{x} | \mathbf{g} \rangle$. On the other hand, photon-mediated interactions consist of long-range dissipative hopping, as the case here, $\Im H_{\text{TB}}=i(H_{\text{TB}}^\dag - H_{\text{TB}})/2$ is no longer diagonal.
Therefore, we consider a generalized spatially resolved escape distribution 
\begin{equation}
F(x,t)=-\int_t^\infty dt'\langle\psi(t')|\{|x\rangle\langle x|,{\rm Im}H_{\text{TB}} \}|\psi(t')\rangle,
\label{eq:escape}
\end{equation}
 which clearly reproduces the conventional definition for on-site loss. The normalization property $\sum_{x} F(x,0) %= -2\int_0^\infty dt'\langle \psi(t')) | {\rm Im}H | \psi(t') \rangle 
 =1$ is shown in Appendix B.
 
In general, $F(x,0)$ is real but not necessarily positive, so it is a quasi-probability distribution, just like the Wigner function \cite{cohen_atom_1998}. The physical intuition behind the negative escape quasi-probability will be explained in the following.

This generalization is motivated from the following decomposition of $d|\langle x|\psi(t)\rangle|^2/dt$:
\begin{equation*}
\begin{split}
\frac{d}{dt}|\langle x|\psi(t)\rangle|^2 &= -i\langle\psi(t)|[|x\rangle\langle x|,{\rm Re}H_{\text{TB}}]|\psi(t)\rangle \\
&+ \langle\psi(t)|\{|x\rangle\langle x|,{\rm Im}H_{\text{TB}}\}|\psi(t)\rangle.
\end{split}
\end{equation*}
The first term on the right-hand side may be interpreted as a coherent current, which persists in the Hermitian limit. The second term is unique to non-Hermitian systems and may be interpreted as incoherent current, including both on-site loss and dissipative hopping from/to other sites. Therefore, due to the non-Hermiticity of the long-range hopping terms, there are incoherent flows between different sites that can lead to net gain at certain sites. 

Returning to the time domain, we know that the short and late-time dynamics generated by $H_{\text{TB}}$ are dominated by radiant and sub-radiant states. The presence of radiant states promotes a rapid exponential fast escape rate in the short-time limit and dilutes the intensity of late-time emission~\cite{reitz_cooperative_2022}. Despite this experimental challenge, we are ideally interested in the spatially resolved late-time dynamics from $t=5$. 

\begin{center}
\begin{figure}[t]
    \centering
    \includegraphics[width=1\columnwidth]{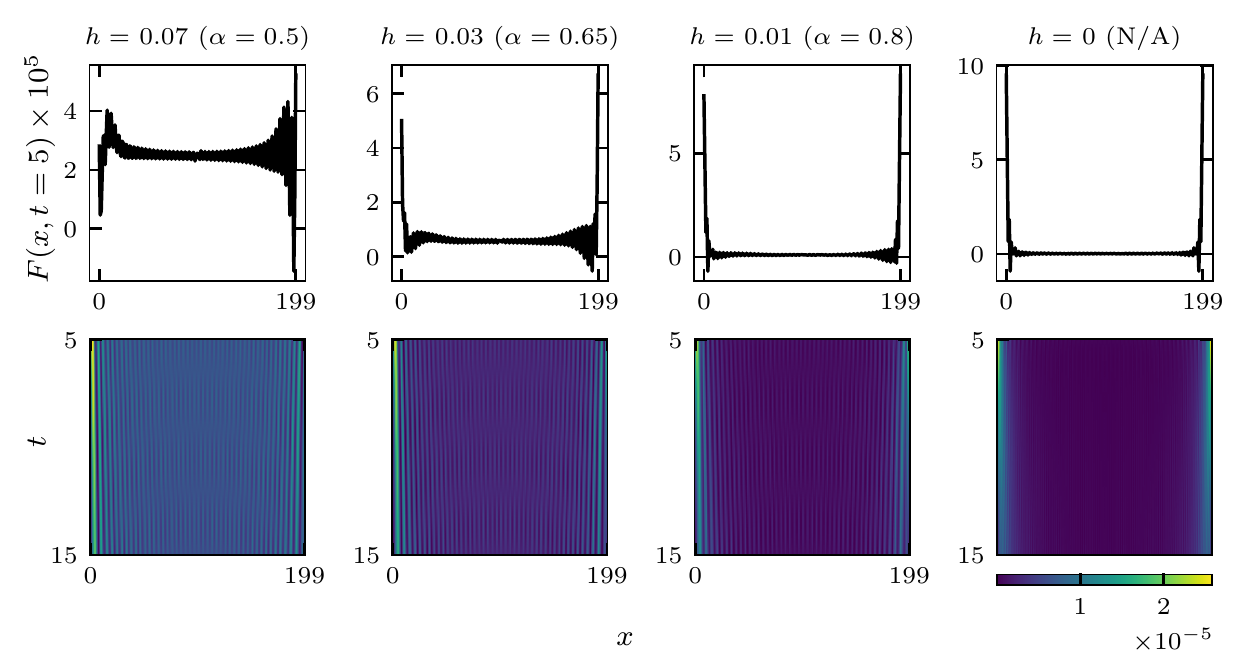}
    \caption{Quantum walk results at four different $h = N^{-\alpha}$ for $N=200$. The RDDI is recovered for $h=0$. The initial state $\left( \sum_{i=1}^{N} \sigma^{\dagger}_{i} \ket{\mathbf{g}} \right) / \sqrt{N}$ is parity symmetric. (Upper) Due to the fast escape rate of radiant states, the short-time limit is excluded. The late-time ($t=5$) escape distribution Eq. (\ref{eq:escape}) shows an asymmetric escape tendency. The emission occurs predominantly at the boundaries of the chain without the edge state. Top left figure shows the existence of the edge state dramatically enhances emission from the bulk. 
    (Lower) A late-time interval wavefunction density, $|\langle x|\psi(t)\rangle|^2$ shows a similar imbalance, where the wavefunction has the greatest overlap with the edge state.}
    \label{fig:quantum_walks}
\end{figure}
\end{center}

Figure \ref{fig:quantum_walks} (Upper) shows the late-time escape distribution $F(x,t=5)$ by propagating the initial state $\ket{\psi_{0}} = \left( \sum_{i=1}^{N} \sigma^{\dagger}_{i} \ket{\mathbf{g}} \right) / \sqrt{N} $, the so-called $W$ state~\cite{dur_three_2000}. The lower panel shows the space-time resolved wavefunction density $|\langle x|\psi(t)\rangle|^2$ from the same initial state under such non-unitary dynamics as a complementary observable. 

For the RDDI model $H_{\text{RDDI}}$ (i.e., $H_{\text{TB}}$ with $h=0$), the excitation is mostly likely to escape from both edges of the atomic chain, which is consistent with the numerical finding of the electric field intensity of the sub-radiant eigenstates in Ref~\cite{asenjo-garcia_exponential_2017}. 

For a small $h$, the winding number of $H_{\text{TB}}$ is not well-defined due to the long-range nature of this model. Hence novel non-Hermitian dynamical phenomena related to bulk topological properties, such as the quantization of walker displacement~\cite{rudner_topological_2009} and edge burst under on-site losses~\cite{xue_non-hermitian_2022} are not observed. However, the existence of an edge state influences the late-time dynamics. For $h < h^{E}_{EP}$, none of the eigenstates are localized, and the excitation is most likely to escape from the boundaries. For $h > h^{E}_{EP}$, we find the disappearance of the boundary concentration and the rise of bulk emission. 

In addition, the escape distribution is asymmetric under parity, while the initial state is symmetric. This is reasonable since the alternating potential explicitly breaks the parity symmetry. This effect is more visible in the time evolution of density --- Fig. \ref{fig:quantum_walks} (lower panels) shows an imbalance across the chain and density accumulation towards the edge state, which reflects that the late-time wavefunction is mostly supported on the edge state. 

\onecolumngrid
\twocolumngrid 

\section{Conclusion}
We have studied the long-range hopping limit of the RDDI Hamiltonian, where the hopping power equals the lattice dimension -- a realistic model for an atomic array in 3D free space. Arranging two species of quantum emitters alternatively, we find the new dispersion relation has a simple form in the infinite chain limit and encodes a novel non-Hermitian behavior, dubbed exceptional-point phase transition. At the transition point, each pair of eigenvectors coalesce, except for two states with the quasi-momentum $k = \pm \pi/(2d)$. 

For a finite chain, the differences in transition energies modify the $N^{-3}$ sub-radiance scaling law and accelerate the most sub-radiant decay rate. The spectrum of finite calculation shows a good agreement with EP phase transition.

We numerically show that one of the $k = \pm \pi/(2d)$ states is boundary localized in real space with a power-law tail for any $h$ greater than the finite size threshold scaling as $N^{-0.65}$, while another is delocalized. The presence of the light line generates discontinuities in the dispersion, which makes the winding number ill-defined. 

Finally, we have generalized the spatial escape distribution for on-site dissipative models to non-local dissipation. For a quantum walker starting from a parity symmetric initial state, both the late-time escape distribution and wavefunction density show spatial imbalances. 

Our work has only focused on 1D arrays in 3D free space. It is natural to consider higher dimensional arrays \cite{perczel_topological_2017}, as well as more general (artificial) photonic environments such as waveguides and photonic crystals that may be engineered to be intrinsically topological or/and dissipative \cite{roccati_hermitian_2023}. It would be interesting to construct more models that exhibit the EP phase transition, to study the topological properties and bulk-edge correspondence, and to explore the rich dynamics of incoherent and collective atomic decay.

\begin{acknowledgments}
    J. Li thanks R. Sch\"afer, D. J. Luitz, F. Piazza, Z. Wang, and L. Piroli for interesting discussions, also Y. E. Zhang for providing technical support for the illustration figure. Z.G. thanks D. Wild for lecturing on the basics of sub-wavelength atomic arrays. J. Li acknowledges support from Deutsche Forschungsgemeinschaft through the project DQUANT (project-id 499347025) and the Erwin Schr\"{o}dinger International Institute for Mathematics and Physics for its hospitality during the Thematic programme Tensor Networks: Mathematical Structures and Novel Algorithms. Z.G. was supported by The University of Tokyo Excellent Young Researcher Program.
\end{acknowledgments}

\typeout{}
\bibliography{ref}

\begin{thebibliography}{63}%
\makeatletter
\providecommand \@ifxundefined [1]{%
 \@ifx{#1\undefined}
}%
\providecommand \@ifnum [1]{%
 \ifnum #1\expandafter \@firstoftwo
 \else \expandafter \@secondoftwo
 \fi
}%
\providecommand \@ifx [1]{%
 \ifx #1\expandafter \@firstoftwo
 \else \expandafter \@secondoftwo
 \fi
}%
\providecommand \natexlab [1]{#1}%
\providecommand \enquote  [1]{``#1''}%
\providecommand \bibnamefont  [1]{#1}%
\providecommand \bibfnamefont [1]{#1}%
\providecommand \citenamefont [1]{#1}%
\providecommand \href@noop [0]{\@secondoftwo}%
\providecommand \href [0]{\begingroup \@sanitize@url \@href}%
\providecommand \@href[1]{\@@startlink{#1}\@@href}%
\providecommand \@@href[1]{\endgroup#1\@@endlink}%
\providecommand \@sanitize@url [0]{\catcode `\\12\catcode `\$12\catcode
  `\&12\catcode `\#12\catcode `\^12\catcode `\_12\catcode `\%12\relax}%
\providecommand \@@startlink[1]{}%
\providecommand \@@endlink[0]{}%
\providecommand \url  [0]{\begingroup\@sanitize@url \@url }%
\providecommand \@url [1]{\endgroup\@href {#1}{\urlprefix }}%
\providecommand \urlprefix  [0]{URL }%
\providecommand \Eprint [0]{\href }%
\providecommand \doibase [0]{https://doi.org/}%
\providecommand \selectlanguage [0]{\@gobble}%
\providecommand \bibinfo  [0]{\@secondoftwo}%
\providecommand \bibfield  [0]{\@secondoftwo}%
\providecommand \translation [1]{[#1]}%
\providecommand \BibitemOpen [0]{}%
\providecommand \bibitemStop [0]{}%
\providecommand \bibitemNoStop [0]{.\EOS\space}%
\providecommand \EOS [0]{\spacefactor3000\relax}%
\providecommand \BibitemShut  [1]{\csname bibitem#1\endcsname}%
\let\auto@bib@innerbib\@empty
%</preamble>
\bibitem [{\citenamefont {Shahmoon}\ \emph {et~al.}(2017)\citenamefont
  {Shahmoon}, \citenamefont {Wild}, \citenamefont {Lukin},\ and\ \citenamefont
  {Yelin}}]{shahmoon_cooperative_2017}%
  \BibitemOpen
  \bibfield  {author} {\bibinfo {author} {\bibfnamefont {E.}~\bibnamefont
  {Shahmoon}}, \bibinfo {author} {\bibfnamefont {D.~S.}\ \bibnamefont {Wild}},
  \bibinfo {author} {\bibfnamefont {M.~D.}\ \bibnamefont {Lukin}},\ and\
  \bibinfo {author} {\bibfnamefont {S.~F.}\ \bibnamefont {Yelin}},\ }\bibfield
  {title} {\bibinfo {title} {Cooperative resonances in light scattering from
  two-dimensional atomic arrays},\ }\href
  {https://doi.org/10.1103/PhysRevLett.118.113601} {\bibfield  {journal}
  {\bibinfo  {journal} {Phys. Rev. Lett.}\ }\textbf {\bibinfo {volume} {118}},\
  \bibinfo {pages} {113601} (\bibinfo {year} {2017})}\BibitemShut {NoStop}%
\bibitem [{\citenamefont {Chang}\ \emph {et~al.}(2018)\citenamefont {Chang},
  \citenamefont {Douglas}, \citenamefont {Gonz\'alez-Tudela}, \citenamefont
  {Hung},\ and\ \citenamefont {Kimble}}]{chang_quantum_2018}%
  \BibitemOpen
  \bibfield  {author} {\bibinfo {author} {\bibfnamefont {D.~E.}\ \bibnamefont
  {Chang}}, \bibinfo {author} {\bibfnamefont {J.~S.}\ \bibnamefont {Douglas}},
  \bibinfo {author} {\bibfnamefont {A.}~\bibnamefont {Gonz\'alez-Tudela}},
  \bibinfo {author} {\bibfnamefont {C.-L.}\ \bibnamefont {Hung}},\ and\
  \bibinfo {author} {\bibfnamefont {H.~J.}\ \bibnamefont {Kimble}},\ }\bibfield
   {title} {\bibinfo {title} {Colloquium: Quantum matter built from nanoscopic
  lattices of atoms and photons},\ }\href
  {https://doi.org/10.1103/RevModPhys.90.031002} {\bibfield  {journal}
  {\bibinfo  {journal} {Rev. Mod. Phys.}\ }\textbf {\bibinfo {volume} {90}},\
  \bibinfo {pages} {031002} (\bibinfo {year} {2018})}\BibitemShut {NoStop}%
\bibitem [{\citenamefont {Manzoni}\ \emph {et~al.}(2018)\citenamefont
  {Manzoni}, \citenamefont {Moreno-Cardoner}, \citenamefont {Asenjo-Garcia},
  \citenamefont {Porto}, \citenamefont {Gorshkov},\ and\ \citenamefont
  {Chang}}]{manzoni_optimization_2018}%
  \BibitemOpen
  \bibfield  {author} {\bibinfo {author} {\bibfnamefont {M.}~\bibnamefont
  {Manzoni}}, \bibinfo {author} {\bibfnamefont {M.}~\bibnamefont
  {Moreno-Cardoner}}, \bibinfo {author} {\bibfnamefont {A.}~\bibnamefont
  {Asenjo-Garcia}}, \bibinfo {author} {\bibfnamefont {J.~V.}\ \bibnamefont
  {Porto}}, \bibinfo {author} {\bibfnamefont {A.~V.}\ \bibnamefont
  {Gorshkov}},\ and\ \bibinfo {author} {\bibfnamefont {D.}~\bibnamefont
  {Chang}},\ }\bibfield  {title} {\bibinfo {title} {Optimization of photon
  storage fidelity in ordered atomic arrays},\ }\href
  {https://iopscience.iop.org/article/10.1088/1367-2630/aadb74/meta} {\bibfield
   {journal} {\bibinfo  {journal} {New J. Phys.}\ }\textbf {\bibinfo {volume}
  {20}},\ \bibinfo {pages} {083048} (\bibinfo {year} {2018})}\BibitemShut
  {NoStop}%
\bibitem [{\citenamefont {Patti}\ \emph {et~al.}(2021)\citenamefont {Patti},
  \citenamefont {Wild}, \citenamefont {Shahmoon}, \citenamefont {Lukin},\ and\
  \citenamefont {Yelin}}]{patti_controlling_2021}%
  \BibitemOpen
  \bibfield  {author} {\bibinfo {author} {\bibfnamefont {T.~L.}\ \bibnamefont
  {Patti}}, \bibinfo {author} {\bibfnamefont {D.~S.}\ \bibnamefont {Wild}},
  \bibinfo {author} {\bibfnamefont {E.}~\bibnamefont {Shahmoon}}, \bibinfo
  {author} {\bibfnamefont {M.~D.}\ \bibnamefont {Lukin}},\ and\ \bibinfo
  {author} {\bibfnamefont {S.~F.}\ \bibnamefont {Yelin}},\ }\bibfield  {title}
  {\bibinfo {title} {Controlling interactions between quantum emitters using
  atom arrays},\ }\href {https://doi.org/10.1103/PhysRevLett.126.223602}
  {\bibfield  {journal} {\bibinfo  {journal} {Phys. Rev. Lett.}\ }\textbf
  {\bibinfo {volume} {126}},\ \bibinfo {pages} {223602} (\bibinfo {year}
  {2021})}\BibitemShut {NoStop}%
\bibitem [{\citenamefont {Moreno-Cardoner}\ \emph {et~al.}(2021)\citenamefont
  {Moreno-Cardoner}, \citenamefont {Goncalves},\ and\ \citenamefont
  {Chang}}]{moreno_quantum_2021}%
  \BibitemOpen
  \bibfield  {author} {\bibinfo {author} {\bibfnamefont {M.}~\bibnamefont
  {Moreno-Cardoner}}, \bibinfo {author} {\bibfnamefont {D.}~\bibnamefont
  {Goncalves}},\ and\ \bibinfo {author} {\bibfnamefont {D.~E.}\ \bibnamefont
  {Chang}},\ }\bibfield  {title} {\bibinfo {title} {Quantum nonlinear optics
  based on two-dimensional rydberg atom arrays},\ }\href
  {https://doi.org/10.1103/PhysRevLett.127.263602} {\bibfield  {journal}
  {\bibinfo  {journal} {Phys. Rev. Lett.}\ }\textbf {\bibinfo {volume} {127}},\
  \bibinfo {pages} {263602} (\bibinfo {year} {2021})}\BibitemShut {NoStop}%
\bibitem [{\citenamefont {Srakaew}\ \emph {et~al.}(2023)\citenamefont
  {Srakaew}, \citenamefont {Weckesser}, \citenamefont {Hollerith},
  \citenamefont {Wei}, \citenamefont {Adler}, \citenamefont {Bloch},\ and\
  \citenamefont {Zeiher}}]{srakaew_a_2023}%
  \BibitemOpen
  \bibfield  {author} {\bibinfo {author} {\bibfnamefont {K.}~\bibnamefont
  {Srakaew}}, \bibinfo {author} {\bibfnamefont {P.}~\bibnamefont {Weckesser}},
  \bibinfo {author} {\bibfnamefont {S.}~\bibnamefont {Hollerith}}, \bibinfo
  {author} {\bibfnamefont {D.}~\bibnamefont {Wei}}, \bibinfo {author}
  {\bibfnamefont {D.}~\bibnamefont {Adler}}, \bibinfo {author} {\bibfnamefont
  {I.}~\bibnamefont {Bloch}},\ and\ \bibinfo {author} {\bibfnamefont
  {J.}~\bibnamefont {Zeiher}},\ }\bibfield  {title} {\bibinfo {title} {A
  subwavelength atomic array switched by a single {Rydberg} atom},\ }\href
  {https://doi.org/10.1038/s41567-023-01959-y} {\bibfield  {journal} {\bibinfo
  {journal} {Nat. Phys.}\ }\textbf {\bibinfo {volume} {19}},\ \bibinfo {pages}
  {714} (\bibinfo {year} {2023})}\BibitemShut {NoStop}%
\bibitem [{\citenamefont {Guerin}\ \emph {et~al.}(2016)\citenamefont {Guerin},
  \citenamefont {Ara\'ujo},\ and\ \citenamefont
  {Kaiser}}]{guerin_subradiance_2016}%
  \BibitemOpen
  \bibfield  {author} {\bibinfo {author} {\bibfnamefont {W.}~\bibnamefont
  {Guerin}}, \bibinfo {author} {\bibfnamefont {M.~O.}\ \bibnamefont
  {Ara\'ujo}},\ and\ \bibinfo {author} {\bibfnamefont {R.}~\bibnamefont
  {Kaiser}},\ }\bibfield  {title} {\bibinfo {title} {Subradiance in a large
  cloud of cold atoms},\ }\href
  {https://doi.org/10.1103/PhysRevLett.116.083601} {\bibfield  {journal}
  {\bibinfo  {journal} {Phys. Rev. Lett.}\ }\textbf {\bibinfo {volume} {116}},\
  \bibinfo {pages} {083601} (\bibinfo {year} {2016})}\BibitemShut {NoStop}%
\bibitem [{\citenamefont {Ferioli}\ \emph {et~al.}(2021)\citenamefont
  {Ferioli}, \citenamefont {Glicenstein}, \citenamefont {Henriet},
  \citenamefont {Ferrier-Barbut},\ and\ \citenamefont
  {Browaeys}}]{ferioli_storage_2021}%
  \BibitemOpen
  \bibfield  {author} {\bibinfo {author} {\bibfnamefont {G.}~\bibnamefont
  {Ferioli}}, \bibinfo {author} {\bibfnamefont {A.}~\bibnamefont
  {Glicenstein}}, \bibinfo {author} {\bibfnamefont {L.}~\bibnamefont
  {Henriet}}, \bibinfo {author} {\bibfnamefont {I.}~\bibnamefont
  {Ferrier-Barbut}},\ and\ \bibinfo {author} {\bibfnamefont {A.}~\bibnamefont
  {Browaeys}},\ }\bibfield  {title} {\bibinfo {title} {Storage and {Release} of
  {Subradiant} {Excitations} in a {Dense} {Atomic} {Cloud}},\ }\href
  {https://doi.org/10.1103/PhysRevX.11.021031} {\bibfield  {journal} {\bibinfo
  {journal} {Phys. Rev. X}\ }\textbf {\bibinfo {volume} {11}},\ \bibinfo
  {pages} {021031} (\bibinfo {year} {2021})}\BibitemShut {NoStop}%
\bibitem [{\citenamefont {Sheremet}\ \emph {et~al.}(2023)\citenamefont
  {Sheremet}, \citenamefont {Petrov}, \citenamefont {Iorsh}, \citenamefont
  {Poshakinskiy},\ and\ \citenamefont {Poddubny}}]{sheremet_waveguide_2023}%
  \BibitemOpen
  \bibfield  {author} {\bibinfo {author} {\bibfnamefont {A.~S.}\ \bibnamefont
  {Sheremet}}, \bibinfo {author} {\bibfnamefont {M.~I.}\ \bibnamefont
  {Petrov}}, \bibinfo {author} {\bibfnamefont {I.~V.}\ \bibnamefont {Iorsh}},
  \bibinfo {author} {\bibfnamefont {A.~V.}\ \bibnamefont {Poshakinskiy}},\ and\
  \bibinfo {author} {\bibfnamefont {A.~N.}\ \bibnamefont {Poddubny}},\
  }\bibfield  {title} {\bibinfo {title} {Waveguide quantum electrodynamics:
  Collective radiance and photon-photon correlations},\ }\href
  {https://doi.org/10.1103/RevModPhys.95.015002} {\bibfield  {journal}
  {\bibinfo  {journal} {Rev. Mod. Phys.}\ }\textbf {\bibinfo {volume} {95}},\
  \bibinfo {pages} {015002} (\bibinfo {year} {2023})}\BibitemShut {NoStop}%
\bibitem [{\citenamefont {Asenjo-Garcia}\ \emph {et~al.}(2017)\citenamefont
  {Asenjo-Garcia}, \citenamefont {Moreno-Cardoner}, \citenamefont {Albrecht},
  \citenamefont {Kimble},\ and\ \citenamefont
  {Chang}}]{asenjo-garcia_exponential_2017}%
  \BibitemOpen
  \bibfield  {author} {\bibinfo {author} {\bibfnamefont {A.}~\bibnamefont
  {Asenjo-Garcia}}, \bibinfo {author} {\bibfnamefont {M.}~\bibnamefont
  {Moreno-Cardoner}}, \bibinfo {author} {\bibfnamefont {A.}~\bibnamefont
  {Albrecht}}, \bibinfo {author} {\bibfnamefont {H.~J.}\ \bibnamefont
  {Kimble}},\ and\ \bibinfo {author} {\bibfnamefont {D.~E.}\ \bibnamefont
  {Chang}},\ }\bibfield  {title} {\bibinfo {title} {Exponential improvement in
  photon storage fidelities using subradiance and "selective radiance" in
  atomic arrays},\ }\href
  {https://doi.org/https://doi.org/10.1103/PhysRevX.7.031024} {\bibfield
  {journal} {\bibinfo  {journal} {Phys. Rev. X}\ }\textbf {\bibinfo {volume}
  {7}},\ \bibinfo {pages} {031024} (\bibinfo {year} {2017})}\BibitemShut
  {NoStop}%
\bibitem [{\citenamefont {Zhang}\ and\ \citenamefont
  {Mølmer}(2019)}]{zhang_theory_2019}%
  \BibitemOpen
  \bibfield  {author} {\bibinfo {author} {\bibfnamefont {Y.-X.}\ \bibnamefont
  {Zhang}}\ and\ \bibinfo {author} {\bibfnamefont {K.}~\bibnamefont
  {Mølmer}},\ }\bibfield  {title} {\bibinfo {title} {Theory of {Subradiant}
  {States} of a {One}-{Dimensional} {Two}-{Level} {Atom} {Chain}},\ }\href
  {https://doi.org/https://doi.org/10.1103/PhysRevLett.122.203605} {\bibfield
  {journal} {\bibinfo  {journal} {Phys. Rev. Lett.}\ }\textbf {\bibinfo
  {volume} {122}},\ \bibinfo {pages} {203605} (\bibinfo {year}
  {2019})}\BibitemShut {NoStop}%
\bibitem [{\citenamefont {Ashida}\ \emph {et~al.}(2020)\citenamefont {Ashida},
  \citenamefont {Gong},\ and\ \citenamefont
  {Ueda}}]{ashida_non-hermitian_2020}%
  \BibitemOpen
  \bibfield  {author} {\bibinfo {author} {\bibfnamefont {Y.}~\bibnamefont
  {Ashida}}, \bibinfo {author} {\bibfnamefont {Z.}~\bibnamefont {Gong}},\ and\
  \bibinfo {author} {\bibfnamefont {M.}~\bibnamefont {Ueda}},\ }\bibfield
  {title} {\bibinfo {title} {Non-{Hermitian} {Physics}},\ }\href
  {https://doi.org/https://doi.org/10.1080/00018732.2021.1876991} {\bibfield
  {journal} {\bibinfo  {journal} {Adv. Phys.}\ }\textbf {\bibinfo {volume}
  {69}},\ \bibinfo {pages} {3} (\bibinfo {year} {2020})}\BibitemShut {NoStop}%
\bibitem [{\citenamefont {Heiss}(2012)}]{heiss_physics_2012}%
  \BibitemOpen
  \bibfield  {author} {\bibinfo {author} {\bibfnamefont {W.~D.}\ \bibnamefont
  {Heiss}},\ }\bibfield  {title} {\bibinfo {title} {The physics of exceptional
  points},\ }\href
  {https://doi.org/https://doi.org/10.1088/1751-8113/45/44/444016} {\bibfield
  {journal} {\bibinfo  {journal} {J. Phys. A Math. Theor.}\ }\textbf {\bibinfo
  {volume} {45}},\ \bibinfo {pages} {444016} (\bibinfo {year}
  {2012})}\BibitemShut {NoStop}%
\bibitem [{\citenamefont {Bergholtz}\ \emph {et~al.}(2021)\citenamefont
  {Bergholtz}, \citenamefont {Budich},\ and\ \citenamefont
  {Kunst}}]{bergholtz_exceptional_2021}%
  \BibitemOpen
  \bibfield  {author} {\bibinfo {author} {\bibfnamefont {E.~J.}\ \bibnamefont
  {Bergholtz}}, \bibinfo {author} {\bibfnamefont {J.~C.}\ \bibnamefont
  {Budich}},\ and\ \bibinfo {author} {\bibfnamefont {F.~K.}\ \bibnamefont
  {Kunst}},\ }\bibfield  {title} {\bibinfo {title} {Exceptional {Topology} of
  {Non}-{Hermitian} {Systems}},\ }\href
  {https://doi.org/10.1103/RevModPhys.93.015005} {\bibfield  {journal}
  {\bibinfo  {journal} {Rev. Mod. Phys.}\ }\textbf {\bibinfo {volume} {93}},\
  \bibinfo {pages} {15005} (\bibinfo {year} {2021})}\BibitemShut {NoStop}%
\bibitem [{\citenamefont {Yao}\ and\ \citenamefont
  {Wang}(2018)}]{yao_edge_2018}%
  \BibitemOpen
  \bibfield  {author} {\bibinfo {author} {\bibfnamefont {S.}~\bibnamefont
  {Yao}}\ and\ \bibinfo {author} {\bibfnamefont {Z.}~\bibnamefont {Wang}},\
  }\bibfield  {title} {\bibinfo {title} {Edge states and topological invariants
  of non-{Hermitian} systems},\ }\href
  {https://doi.org/https://doi.org/10.1103/PhysRevLett.121.086803} {\bibfield
  {journal} {\bibinfo  {journal} {Phys. Rev. Lett.}\ }\textbf {\bibinfo
  {volume} {121}},\ \bibinfo {pages} {086803} (\bibinfo {year}
  {2018})}\BibitemShut {NoStop}%
\bibitem [{\citenamefont {Okuma}\ and\ \citenamefont {Sato}(2023)}]{Okuma2023}%
  \BibitemOpen
  \bibfield  {author} {\bibinfo {author} {\bibfnamefont {N.}~\bibnamefont
  {Okuma}}\ and\ \bibinfo {author} {\bibfnamefont {M.}~\bibnamefont {Sato}},\
  }\bibfield  {title} {\bibinfo {title} {Non-hermitian topological phenomena: A
  review},\ }\href {https://doi.org/10.1146/annurev-conmatphys-040521-033133}
  {\bibfield  {journal} {\bibinfo  {journal} {Annu. Rev. Condens. Matter
  Phys.}\ }\textbf {\bibinfo {volume} {14}},\ \bibinfo {pages} {83} (\bibinfo
  {year} {2023})}\BibitemShut {NoStop}%
\bibitem [{\citenamefont {Gong}\ \emph {et~al.}(2018)\citenamefont {Gong},
  \citenamefont {Ashida}, \citenamefont {Kawabata}, \citenamefont {Takasan},
  \citenamefont {Higashikawa},\ and\ \citenamefont
  {Ueda}}]{gong_topological_2018}%
  \BibitemOpen
  \bibfield  {author} {\bibinfo {author} {\bibfnamefont {Z.}~\bibnamefont
  {Gong}}, \bibinfo {author} {\bibfnamefont {Y.}~\bibnamefont {Ashida}},
  \bibinfo {author} {\bibfnamefont {K.}~\bibnamefont {Kawabata}}, \bibinfo
  {author} {\bibfnamefont {K.}~\bibnamefont {Takasan}}, \bibinfo {author}
  {\bibfnamefont {S.}~\bibnamefont {Higashikawa}},\ and\ \bibinfo {author}
  {\bibfnamefont {M.}~\bibnamefont {Ueda}},\ }\bibfield  {title} {\bibinfo
  {title} {Topological phases of non-{Hermitian} systems},\ }\href
  {https://doi.org/10.1103/PhysRevX.8.031079} {\bibfield  {journal} {\bibinfo
  {journal} {Phys. Rev. X}\ }\textbf {\bibinfo {volume} {8}},\ \bibinfo {pages}
  {031079} (\bibinfo {year} {2018})}\BibitemShut {NoStop}%
\bibitem [{\citenamefont {Kawabata}\ \emph
  {et~al.}(2019{\natexlab{a}})\citenamefont {Kawabata}, \citenamefont
  {Shiozaki}, \citenamefont {Ueda},\ and\ \citenamefont
  {Sato}}]{kawabata_symmetry_2019}%
  \BibitemOpen
  \bibfield  {author} {\bibinfo {author} {\bibfnamefont {K.}~\bibnamefont
  {Kawabata}}, \bibinfo {author} {\bibfnamefont {K.}~\bibnamefont {Shiozaki}},
  \bibinfo {author} {\bibfnamefont {M.}~\bibnamefont {Ueda}},\ and\ \bibinfo
  {author} {\bibfnamefont {M.}~\bibnamefont {Sato}},\ }\bibfield  {title}
  {\bibinfo {title} {Symmetry and {Topology} in {Non}-{Hermitian} {Physics}},\
  }\href {https://doi.org/https://doi.org/10.1103/PhysRevX.9.041015} {\bibfield
   {journal} {\bibinfo  {journal} {Phys. Rev. X}\ }\textbf {\bibinfo {volume}
  {9}},\ \bibinfo {pages} {041015} (\bibinfo {year}
  {2019}{\natexlab{a}})}\BibitemShut {NoStop}%
\bibitem [{\citenamefont {Zhou}\ and\ \citenamefont {Lee}(2019)}]{Zhou2019}%
  \BibitemOpen
  \bibfield  {author} {\bibinfo {author} {\bibfnamefont {H.}~\bibnamefont
  {Zhou}}\ and\ \bibinfo {author} {\bibfnamefont {J.~Y.}\ \bibnamefont {Lee}},\
  }\bibfield  {title} {\bibinfo {title} {Periodic table for topological bands
  with non-hermitian symmetries},\ }\href
  {https://doi.org/10.1103/PhysRevB.99.235112} {\bibfield  {journal} {\bibinfo
  {journal} {Phys. Rev. B}\ }\textbf {\bibinfo {volume} {99}},\ \bibinfo
  {pages} {235112} (\bibinfo {year} {2019})}\BibitemShut {NoStop}%
\bibitem [{\citenamefont {Gong}\ \emph
  {et~al.}(2022{\natexlab{a}})\citenamefont {Gong}, \citenamefont {Bello},
  \citenamefont {Malz},\ and\ \citenamefont {Kunst}}]{gong_anomalous_2022}%
  \BibitemOpen
  \bibfield  {author} {\bibinfo {author} {\bibfnamefont {Z.}~\bibnamefont
  {Gong}}, \bibinfo {author} {\bibfnamefont {M.}~\bibnamefont {Bello}},
  \bibinfo {author} {\bibfnamefont {D.}~\bibnamefont {Malz}},\ and\ \bibinfo
  {author} {\bibfnamefont {F.~K.}\ \bibnamefont {Kunst}},\ }\bibfield  {title}
  {\bibinfo {title} {Anomalous {Behaviors} of {Quantum} {Emitters} in
  {Non}-{Hermitian} {Baths}},\ }\href
  {https://doi.org/10.1103/PhysRevLett.129.223601} {\bibfield  {journal}
  {\bibinfo  {journal} {Phys. Rev. Lett.}\ }\textbf {\bibinfo {volume} {129}},\
  \bibinfo {pages} {223601} (\bibinfo {year} {2022}{\natexlab{a}})}\BibitemShut
  {NoStop}%
\bibitem [{\citenamefont {Gong}\ \emph
  {et~al.}(2022{\natexlab{b}})\citenamefont {Gong}, \citenamefont {Bello},
  \citenamefont {Malz},\ and\ \citenamefont {Kunst}}]{gong_bound_2022}%
  \BibitemOpen
  \bibfield  {author} {\bibinfo {author} {\bibfnamefont {Z.}~\bibnamefont
  {Gong}}, \bibinfo {author} {\bibfnamefont {M.}~\bibnamefont {Bello}},
  \bibinfo {author} {\bibfnamefont {D.}~\bibnamefont {Malz}},\ and\ \bibinfo
  {author} {\bibfnamefont {F.~K.}\ \bibnamefont {Kunst}},\ }\bibfield  {title}
  {\bibinfo {title} {Bound states and photon emission in non-{Hermitian}
  nanophotonics},\ }\href {https://doi.org/10.1103/PhysRevA.106.053517}
  {\bibfield  {journal} {\bibinfo  {journal} {Phys. Rev. A}\ }\textbf {\bibinfo
  {volume} {106}},\ \bibinfo {pages} {053517} (\bibinfo {year}
  {2022}{\natexlab{b}})}\BibitemShut {NoStop}%
\bibitem [{\citenamefont {Roccati}\ \emph {et~al.}()\citenamefont {Roccati},
  \citenamefont {Bello}, \citenamefont {Gong}, \citenamefont {Ueda},
  \citenamefont {Ciccarello}, \citenamefont {Chenu},\ and\ \citenamefont
  {Carollo}}]{roccati_hermitian_2023}%
  \BibitemOpen
  \bibfield  {author} {\bibinfo {author} {\bibfnamefont {F.}~\bibnamefont
  {Roccati}}, \bibinfo {author} {\bibfnamefont {M.}~\bibnamefont {Bello}},
  \bibinfo {author} {\bibfnamefont {Z.}~\bibnamefont {Gong}}, \bibinfo {author}
  {\bibfnamefont {M.}~\bibnamefont {Ueda}}, \bibinfo {author} {\bibfnamefont
  {F.}~\bibnamefont {Ciccarello}}, \bibinfo {author} {\bibfnamefont
  {A.}~\bibnamefont {Chenu}},\ and\ \bibinfo {author} {\bibfnamefont
  {A.}~\bibnamefont {Carollo}},\ }\bibfield  {title} {\bibinfo {title}
  {Hermitian and non-hermitian topology from photon-mediated interactions},\
  }\href@noop {} {\ }\Eprint {https://arxiv.org/abs/2303.00762}
  {arXiv:2303.00762} \BibitemShut {NoStop}%
\bibitem [{\citenamefont {Xiao}\ \emph {et~al.}(2020)\citenamefont {Xiao},
  \citenamefont {Deng}, \citenamefont {Wang}, \citenamefont {Zhu},
  \citenamefont {Wang}, \citenamefont {Yi},\ and\ \citenamefont
  {Xue}}]{xiao_observation_2020}%
  \BibitemOpen
  \bibfield  {author} {\bibinfo {author} {\bibfnamefont {L.}~\bibnamefont
  {Xiao}}, \bibinfo {author} {\bibfnamefont {T.}~\bibnamefont {Deng}}, \bibinfo
  {author} {\bibfnamefont {K.}~\bibnamefont {Wang}}, \bibinfo {author}
  {\bibfnamefont {G.}~\bibnamefont {Zhu}}, \bibinfo {author} {\bibfnamefont
  {Z.}~\bibnamefont {Wang}}, \bibinfo {author} {\bibfnamefont {W.}~\bibnamefont
  {Yi}},\ and\ \bibinfo {author} {\bibfnamefont {P.}~\bibnamefont {Xue}},\
  }\bibfield  {title} {\bibinfo {title} {Observation of non-{Hermitian}
  bulk-boundary correspondence in quantum dynamics},\ }\href
  {https://doi.org/https://doi.org/10.1038/s41567-020-0836-6} {\bibfield
  {journal} {\bibinfo  {journal} {Nat. Phys.}\ }\textbf {\bibinfo {volume}
  {16}},\ \bibinfo {pages} {761} (\bibinfo {year} {2020})}\BibitemShut
  {NoStop}%
\bibitem [{\citenamefont {Xiao}\ \emph {et~al.}()\citenamefont {Xiao},
  \citenamefont {Xue}, \citenamefont {Song}, \citenamefont {Hu}, \citenamefont
  {Yi}, \citenamefont {Wang},\ and\ \citenamefont
  {Xue}}]{xiao_observation_2023}%
  \BibitemOpen
  \bibfield  {author} {\bibinfo {author} {\bibfnamefont {L.}~\bibnamefont
  {Xiao}}, \bibinfo {author} {\bibfnamefont {W.-T.}\ \bibnamefont {Xue}},
  \bibinfo {author} {\bibfnamefont {F.}~\bibnamefont {Song}}, \bibinfo {author}
  {\bibfnamefont {Y.-M.}\ \bibnamefont {Hu}}, \bibinfo {author} {\bibfnamefont
  {W.}~\bibnamefont {Yi}}, \bibinfo {author} {\bibfnamefont {Z.}~\bibnamefont
  {Wang}},\ and\ \bibinfo {author} {\bibfnamefont {P.}~\bibnamefont {Xue}},\
  }\bibfield  {title} {\bibinfo {title} {Observation of non-hermitian edge
  burst in quantum dynamics},\ }\href@noop {} {\ }\Eprint
  {https://arxiv.org/abs/2303.12831} {arXiv:2303.12831} \BibitemShut {NoStop}%
\bibitem [{\citenamefont {Kuwahara}\ and\ \citenamefont
  {Saito}(2021)}]{Kuwahara2021}%
  \BibitemOpen
  \bibfield  {author} {\bibinfo {author} {\bibfnamefont {T.}~\bibnamefont
  {Kuwahara}}\ and\ \bibinfo {author} {\bibfnamefont {K.}~\bibnamefont
  {Saito}},\ }\bibfield  {title} {\bibinfo {title} {Absence of fast scrambling
  in thermodynamically stable long-range interacting systems},\ }\href
  {https://doi.org/10.1103/PhysRevLett.126.030604} {\bibfield  {journal}
  {\bibinfo  {journal} {Phys. Rev. Lett.}\ }\textbf {\bibinfo {volume} {126}},\
  \bibinfo {pages} {030604} (\bibinfo {year} {2021})}\BibitemShut {NoStop}%
\bibitem [{\citenamefont {Gong}\ \emph {et~al.}(2023)\citenamefont {Gong},
  \citenamefont {Guaita},\ and\ \citenamefont {Cirac}}]{gong_long-range_2023}%
  \BibitemOpen
  \bibfield  {author} {\bibinfo {author} {\bibfnamefont {Z.}~\bibnamefont
  {Gong}}, \bibinfo {author} {\bibfnamefont {T.}~\bibnamefont {Guaita}},\ and\
  \bibinfo {author} {\bibfnamefont {J.~I.}\ \bibnamefont {Cirac}},\ }\bibfield
  {title} {\bibinfo {title} {Long-{Range} {Free} {Fermions}: {Lieb}-{Robinson}
  {Bound}, {Clustering} {Properties}, and {Topological} {Phases}},\ }\href
  {https://doi.org/10.1103/PhysRevLett.130.070401} {\bibfield  {journal}
  {\bibinfo  {journal} {Phys. Rev. Lett.}\ }\textbf {\bibinfo {volume} {130}},\
  \bibinfo {pages} {070401} (\bibinfo {year} {2023})}\BibitemShut {NoStop}%
\bibitem [{\citenamefont {Lehmberg}(1970)}]{lehmberg_radiation_1970}%
  \BibitemOpen
  \bibfield  {author} {\bibinfo {author} {\bibfnamefont {R.~H.}\ \bibnamefont
  {Lehmberg}},\ }\bibfield  {title} {\bibinfo {title} {Radiation from an
  $n$-atom system. i. general formalism},\ }\href
  {https://doi.org/10.1103/PhysRevA.2.883} {\bibfield  {journal} {\bibinfo
  {journal} {Phys. Rev. A}\ }\textbf {\bibinfo {volume} {2}},\ \bibinfo {pages}
  {883} (\bibinfo {year} {1970})}\BibitemShut {NoStop}%
\bibitem [{\citenamefont {Gruner}\ and\ \citenamefont
  {Welsch}(1996)}]{gruner_green-function_1996}%
  \BibitemOpen
  \bibfield  {author} {\bibinfo {author} {\bibfnamefont {T.}~\bibnamefont
  {Gruner}}\ and\ \bibinfo {author} {\bibfnamefont {D.-G.}\ \bibnamefont
  {Welsch}},\ }\bibfield  {title} {\bibinfo {title} {Green-function approach to
  the radiation-field quantization for homogeneous and inhomogeneous
  {Kramers}-{Kronig} dielectrics},\ }\href
  {https://doi.org/10.1103/PhysRevA.53.1818} {\bibfield  {journal} {\bibinfo
  {journal} {Phys. Rev. A}\ }\textbf {\bibinfo {volume} {53}},\ \bibinfo
  {pages} {1818} (\bibinfo {year} {1996})}\BibitemShut {NoStop}%
\bibitem [{\citenamefont {Xu}\ and\ \citenamefont
  {Fan}(2015)}]{xu_input-output_2015}%
  \BibitemOpen
  \bibfield  {author} {\bibinfo {author} {\bibfnamefont {S.}~\bibnamefont
  {Xu}}\ and\ \bibinfo {author} {\bibfnamefont {S.}~\bibnamefont {Fan}},\
  }\bibfield  {title} {\bibinfo {title} {Input-output formalism for few-photon
  transport: A systematic treatment beyond two photons},\ }\href
  {https://doi.org/10.1103/PhysRevA.91.043845} {\bibfield  {journal} {\bibinfo
  {journal} {Phys. Rev. A}\ }\textbf {\bibinfo {volume} {91}},\ \bibinfo
  {pages} {043845} (\bibinfo {year} {2015})}\BibitemShut {NoStop}%
\bibitem [{\citenamefont {Shi}\ \emph {et~al.}(2015)\citenamefont {Shi},
  \citenamefont {Chang},\ and\ \citenamefont {Cirac}}]{shi_multi-photon_2015}%
  \BibitemOpen
  \bibfield  {author} {\bibinfo {author} {\bibfnamefont {T.}~\bibnamefont
  {Shi}}, \bibinfo {author} {\bibfnamefont {D.~E.}\ \bibnamefont {Chang}},\
  and\ \bibinfo {author} {\bibfnamefont {J.~I.}\ \bibnamefont {Cirac}},\
  }\bibfield  {title} {\bibinfo {title} {Multi-photon {Scattering} {Theory} and
  {Generalized} {Master} {Equations}},\ }\href
  {https://doi.org/https://doi.org/10.1103/PhysRevA.92.053834} {\bibfield
  {journal} {\bibinfo  {journal} {Phys. Rev. A}\ }\textbf {\bibinfo {volume}
  {92}},\ \bibinfo {pages} {053834} (\bibinfo {year} {2015})}\BibitemShut
  {NoStop}%
\bibitem [{\citenamefont {Zhang}\ and\ \citenamefont
  {Mølmer}(2020)}]{zhang_subradiant_2020-1}%
  \BibitemOpen
  \bibfield  {author} {\bibinfo {author} {\bibfnamefont {Y.-X.}\ \bibnamefont
  {Zhang}}\ and\ \bibinfo {author} {\bibfnamefont {K.}~\bibnamefont
  {Mølmer}},\ }\bibfield  {title} {\bibinfo {title} {Subradiant emission from
  regular atomic arrays: universal scaling of decay rates from the generalized
  {Bloch} theorem},\ }\href
  {https://doi.org/https://doi.org/10.1103/PhysRevLett.125.253601} {\bibfield
  {journal} {\bibinfo  {journal} {Phys. Rev. Lett.}\ }\textbf {\bibinfo
  {volume} {125}},\ \bibinfo {pages} {253601} (\bibinfo {year}
  {2020})}\BibitemShut {NoStop}%
\bibitem [{\citenamefont {Reitz}\ \emph {et~al.}(2022)\citenamefont {Reitz},
  \citenamefont {Sommer},\ and\ \citenamefont
  {Genes}}]{reitz_cooperative_2022}%
  \BibitemOpen
  \bibfield  {author} {\bibinfo {author} {\bibfnamefont {M.}~\bibnamefont
  {Reitz}}, \bibinfo {author} {\bibfnamefont {C.}~\bibnamefont {Sommer}},\ and\
  \bibinfo {author} {\bibfnamefont {C.}~\bibnamefont {Genes}},\ }\bibfield
  {title} {\bibinfo {title} {Cooperative {Quantum} {Phenomena} in
  {Light}-{Matter} {Platforms}},\ }\href
  {https://doi.org/10.1103/PRXQuantum.3.010201} {\bibfield  {journal} {\bibinfo
   {journal} {PRX Quantum}\ }\textbf {\bibinfo {volume} {3}},\ \bibinfo {pages}
  {010201} (\bibinfo {year} {2022})}\BibitemShut {NoStop}%
\bibitem [{\citenamefont {Harper}(1955)}]{Harper1955}%
  \BibitemOpen
  \bibfield  {author} {\bibinfo {author} {\bibfnamefont {P.~G.}\ \bibnamefont
  {Harper}},\ }\bibfield  {title} {\bibinfo {title} {Single band motion of
  conduction electrons in a uniform magnetic field},\ }\href
  {https://doi.org/10.1088/0370-1298/68/10/304} {\bibfield  {journal} {\bibinfo
   {journal} {Proc. Phys. Soc. Lond. A}\ }\textbf {\bibinfo {volume} {68}},\
  \bibinfo {pages} {874} (\bibinfo {year} {1955})}\BibitemShut {NoStop}%
\bibitem [{\citenamefont {Aubry}\ and\ \citenamefont
  {Andr\'e}(1980)}]{Aubry1980}%
  \BibitemOpen
  \bibfield  {author} {\bibinfo {author} {\bibfnamefont {S.}~\bibnamefont
  {Aubry}}\ and\ \bibinfo {author} {\bibfnamefont {G.}~\bibnamefont
  {Andr\'e}},\ }\bibfield  {title} {\bibinfo {title} {Analyticity breaking and
  anderson localization in incommensurate lattices},\ }\href@noop {} {\bibfield
   {journal} {\bibinfo  {journal} {Ann. Israel Phys. Soc.}\ }\textbf {\bibinfo
  {volume} {3}},\ \bibinfo {pages} {18} (\bibinfo {year} {1980})}\BibitemShut
  {NoStop}%
\bibitem [{\citenamefont {Tzortzakakis}\ \emph {et~al.}(2022)\citenamefont
  {Tzortzakakis}, \citenamefont {Katsaris}, \citenamefont {Palaiodimopoulos},
  \citenamefont {Kalozoumis}, \citenamefont {Theocharis}, \citenamefont
  {Diakonos},\ and\ \citenamefont {Petrosyan}}]{tzortzakakis_topological_2022}%
  \BibitemOpen
  \bibfield  {author} {\bibinfo {author} {\bibfnamefont {A.~F.}\ \bibnamefont
  {Tzortzakakis}}, \bibinfo {author} {\bibfnamefont {A.}~\bibnamefont
  {Katsaris}}, \bibinfo {author} {\bibfnamefont {N.~E.}\ \bibnamefont
  {Palaiodimopoulos}}, \bibinfo {author} {\bibfnamefont {P.~A.}\ \bibnamefont
  {Kalozoumis}}, \bibinfo {author} {\bibfnamefont {G.}~\bibnamefont
  {Theocharis}}, \bibinfo {author} {\bibfnamefont {F.~K.}\ \bibnamefont
  {Diakonos}},\ and\ \bibinfo {author} {\bibfnamefont {D.}~\bibnamefont
  {Petrosyan}},\ }\bibfield  {title} {\bibinfo {title} {Topological edge states
  of the $\mathcal{PT}$-symmetric su-schrieffer-heeger model: An effective
  two-state description},\ }\href {https://doi.org/10.1103/PhysRevA.106.023513}
  {\bibfield  {journal} {\bibinfo  {journal} {Phys. Rev. A}\ }\textbf {\bibinfo
  {volume} {106}},\ \bibinfo {pages} {023513} (\bibinfo {year}
  {2022})}\BibitemShut {NoStop}%
\bibitem [{\citenamefont {Weimann}\ \emph {et~al.}(2017)\citenamefont
  {Weimann}, \citenamefont {Kremer}, \citenamefont {Plotnik}, \citenamefont
  {Lumer}, \citenamefont {Nolte}, \citenamefont {Makris}, \citenamefont
  {Segev}, \citenamefont {Rechtsman},\ and\ \citenamefont
  {Szameit}}]{weimann_topologically_2017}%
  \BibitemOpen
  \bibfield  {author} {\bibinfo {author} {\bibfnamefont {S.}~\bibnamefont
  {Weimann}}, \bibinfo {author} {\bibfnamefont {M.}~\bibnamefont {Kremer}},
  \bibinfo {author} {\bibfnamefont {Y.}~\bibnamefont {Plotnik}}, \bibinfo
  {author} {\bibfnamefont {Y.}~\bibnamefont {Lumer}}, \bibinfo {author}
  {\bibfnamefont {S.}~\bibnamefont {Nolte}}, \bibinfo {author} {\bibfnamefont
  {K.~G.}\ \bibnamefont {Makris}}, \bibinfo {author} {\bibfnamefont
  {M.}~\bibnamefont {Segev}}, \bibinfo {author} {\bibfnamefont {M.~C.}\
  \bibnamefont {Rechtsman}},\ and\ \bibinfo {author} {\bibfnamefont
  {A.}~\bibnamefont {Szameit}},\ }\bibfield  {title} {\bibinfo {title}
  {Topologically protected bound states in photonic parity--time-symmetric
  crystals},\ }\href {https://www.nature.com/articles/nmat4811} {\bibfield
  {journal} {\bibinfo  {journal} {Nat. Mater.}\ }\textbf {\bibinfo {volume}
  {16}},\ \bibinfo {pages} {433} (\bibinfo {year} {2017})}\BibitemShut
  {NoStop}%
\bibitem [{\citenamefont {Biddle}\ and\ \citenamefont
  {Das~Sarma}(2010)}]{biddle_predicted_2010}%
  \BibitemOpen
  \bibfield  {author} {\bibinfo {author} {\bibfnamefont {J.}~\bibnamefont
  {Biddle}}\ and\ \bibinfo {author} {\bibfnamefont {S.}~\bibnamefont
  {Das~Sarma}},\ }\bibfield  {title} {\bibinfo {title} {Predicted mobility
  edges in one-dimensional incommensurate optical lattices: An exactly solvable
  model of anderson localization},\ }\href
  {https://doi.org/10.1103/PhysRevLett.104.070601} {\bibfield  {journal}
  {\bibinfo  {journal} {Phys. Rev. Lett.}\ }\textbf {\bibinfo {volume} {104}},\
  \bibinfo {pages} {070601} (\bibinfo {year} {2010})}\BibitemShut {NoStop}%
\bibitem [{\citenamefont {Ganeshan}\ \emph {et~al.}(2015)\citenamefont
  {Ganeshan}, \citenamefont {Pixley},\ and\ \citenamefont
  {Das~Sarma}}]{ganeshan_nearest_2015}%
  \BibitemOpen
  \bibfield  {author} {\bibinfo {author} {\bibfnamefont {S.}~\bibnamefont
  {Ganeshan}}, \bibinfo {author} {\bibfnamefont {J.~H.}\ \bibnamefont
  {Pixley}},\ and\ \bibinfo {author} {\bibfnamefont {S.}~\bibnamefont
  {Das~Sarma}},\ }\bibfield  {title} {\bibinfo {title} {Nearest neighbor tight
  binding models with an exact mobility edge in one dimension},\ }\href
  {https://doi.org/10.1103/PhysRevLett.114.146601} {\bibfield  {journal}
  {\bibinfo  {journal} {Phys. Rev. Lett.}\ }\textbf {\bibinfo {volume} {114}},\
  \bibinfo {pages} {146601} (\bibinfo {year} {2015})}\BibitemShut {NoStop}%
\bibitem [{\citenamefont {Wang}\ \emph {et~al.}(2020)\citenamefont {Wang},
  \citenamefont {Xia}, \citenamefont {Zhang}, \citenamefont {Yao},
  \citenamefont {Chen}, \citenamefont {You}, \citenamefont {Zhou},\ and\
  \citenamefont {Liu}}]{Liu2020}%
  \BibitemOpen
  \bibfield  {author} {\bibinfo {author} {\bibfnamefont {Y.}~\bibnamefont
  {Wang}}, \bibinfo {author} {\bibfnamefont {X.}~\bibnamefont {Xia}}, \bibinfo
  {author} {\bibfnamefont {L.}~\bibnamefont {Zhang}}, \bibinfo {author}
  {\bibfnamefont {H.}~\bibnamefont {Yao}}, \bibinfo {author} {\bibfnamefont
  {S.}~\bibnamefont {Chen}}, \bibinfo {author} {\bibfnamefont {J.}~\bibnamefont
  {You}}, \bibinfo {author} {\bibfnamefont {Q.}~\bibnamefont {Zhou}},\ and\
  \bibinfo {author} {\bibfnamefont {X.-J.}\ \bibnamefont {Liu}},\ }\bibfield
  {title} {\bibinfo {title} {One-dimensional quasiperiodic mosaic lattice with
  exact mobility edges},\ }\href
  {https://doi.org/10.1103/PhysRevLett.125.196604} {\bibfield  {journal}
  {\bibinfo  {journal} {Phys. Rev. Lett.}\ }\textbf {\bibinfo {volume} {125}},\
  \bibinfo {pages} {196604} (\bibinfo {year} {2020})}\BibitemShut {NoStop}%
\bibitem [{\citenamefont {Luitz}\ and\ \citenamefont
  {Piazza}(2019)}]{luitz_exceptional_2019}%
  \BibitemOpen
  \bibfield  {author} {\bibinfo {author} {\bibfnamefont {D.~J.}\ \bibnamefont
  {Luitz}}\ and\ \bibinfo {author} {\bibfnamefont {F.}~\bibnamefont {Piazza}},\
  }\bibfield  {title} {\bibinfo {title} {Exceptional points and the topology of
  quantum many-body spectra},\ }\href
  {https://doi.org/10.1103/PhysRevResearch.1.033051} {\bibfield  {journal}
  {\bibinfo  {journal} {Phys. Rev. Research}\ }\textbf {\bibinfo {volume}
  {1}},\ \bibinfo {pages} {033051} (\bibinfo {year} {2019})}\BibitemShut
  {NoStop}%
\bibitem [{\citenamefont {Schäfer}\ \emph {et~al.}(2022)\citenamefont
  {Schäfer}, \citenamefont {Budich},\ and\ \citenamefont
  {Luitz}}]{schafer_symmetry_2022}%
  \BibitemOpen
  \bibfield  {author} {\bibinfo {author} {\bibfnamefont {R.}~\bibnamefont
  {Schäfer}}, \bibinfo {author} {\bibfnamefont {J.~C.}\ \bibnamefont
  {Budich}},\ and\ \bibinfo {author} {\bibfnamefont {D.~J.}\ \bibnamefont
  {Luitz}},\ }\bibfield  {title} {\bibinfo {title} {Symmetry protected
  exceptional points of interacting fermions},\ }\href
  {https://doi.org/10.1103/PhysRevResearch.4.033181} {\bibfield  {journal}
  {\bibinfo  {journal} {Phys. Rev. Research}\ }\textbf {\bibinfo {volume}
  {4}},\ \bibinfo {pages} {033181} (\bibinfo {year} {2022})}\BibitemShut
  {NoStop}%
\bibitem [{\citenamefont {Riley}\ \emph {et~al.}(2006)\citenamefont {Riley},
  \citenamefont {Hobson},\ and\ \citenamefont
  {Bence}}]{riley_mathematical_2006}%
  \BibitemOpen
  \bibfield  {author} {\bibinfo {author} {\bibfnamefont {K.~F.}\ \bibnamefont
  {Riley}}, \bibinfo {author} {\bibfnamefont {M.~P.}\ \bibnamefont {Hobson}},\
  and\ \bibinfo {author} {\bibfnamefont {S.~J.}\ \bibnamefont {Bence}},\
  }\href@noop {} {\emph {\bibinfo {title} {Mathematical Methods for Physics and
  Engineering: A Comprehensive Guide}}}\ (\bibinfo  {publisher} {Cambridge
  University Press},\ \bibinfo {year} {2006})\BibitemShut {NoStop}%
\bibitem [{\citenamefont {Kornovan}\ \emph {et~al.}(2019)\citenamefont
  {Kornovan}, \citenamefont {Corzo}, \citenamefont {Laurat},\ and\
  \citenamefont {Sheremet}}]{kornovan_extremely_2019}%
  \BibitemOpen
  \bibfield  {author} {\bibinfo {author} {\bibfnamefont {D.~F.}\ \bibnamefont
  {Kornovan}}, \bibinfo {author} {\bibfnamefont {N.~V.}\ \bibnamefont {Corzo}},
  \bibinfo {author} {\bibfnamefont {J.}~\bibnamefont {Laurat}},\ and\ \bibinfo
  {author} {\bibfnamefont {A.~S.}\ \bibnamefont {Sheremet}},\ }\bibfield
  {title} {\bibinfo {title} {Extremely subradiant states in a periodic
  one-dimensional atomic array},\ }\href
  {https://doi.org/10.1103/PhysRevA.100.063832} {\bibfield  {journal} {\bibinfo
   {journal} {Phys. Rev. A}\ }\textbf {\bibinfo {volume} {100}},\ \bibinfo
  {pages} {063832} (\bibinfo {year} {2019})}\BibitemShut {NoStop}%
\bibitem [{\citenamefont {Vodola}\ \emph {et~al.}(2015)\citenamefont {Vodola},
  \citenamefont {Lepori}, \citenamefont {Ercolessi},\ and\ \citenamefont
  {Pupillo}}]{Vodola_long-range_2016}%
  \BibitemOpen
  \bibfield  {author} {\bibinfo {author} {\bibfnamefont {D.}~\bibnamefont
  {Vodola}}, \bibinfo {author} {\bibfnamefont {L.}~\bibnamefont {Lepori}},
  \bibinfo {author} {\bibfnamefont {E.}~\bibnamefont {Ercolessi}},\ and\
  \bibinfo {author} {\bibfnamefont {G.}~\bibnamefont {Pupillo}},\ }\bibfield
  {title} {\bibinfo {title} {Long-range ising and kitaev models: phases,
  correlations and edge modes},\ }\href
  {https://doi.org/10.1088/1367-2630/18/1/015001} {\bibfield  {journal}
  {\bibinfo  {journal} {New J. Phys.}\ }\textbf {\bibinfo {volume} {18}},\
  \bibinfo {pages} {015001} (\bibinfo {year} {2015})}\BibitemShut {NoStop}%
\bibitem [{\citenamefont {Jäger}\ \emph {et~al.}(2020)\citenamefont {Jäger},
  \citenamefont {Dell'Anna},\ and\ \citenamefont {Morigi}}]{jager_edge_2020}%
  \BibitemOpen
  \bibfield  {author} {\bibinfo {author} {\bibfnamefont {S.~B.}\ \bibnamefont
  {Jäger}}, \bibinfo {author} {\bibfnamefont {L.}~\bibnamefont {Dell'Anna}},\
  and\ \bibinfo {author} {\bibfnamefont {G.}~\bibnamefont {Morigi}},\
  }\bibfield  {title} {\bibinfo {title} {Edge states of the long-range {Kitaev}
  chain: an analytical study},\ }\href
  {https://doi.org/https://doi.org/10.1103/PhysRevB.102.035152} {\bibfield
  {journal} {\bibinfo  {journal} {Phys. Rev. B}\ }\textbf {\bibinfo {volume}
  {102}},\ \bibinfo {pages} {035152} (\bibinfo {year} {2020})}\BibitemShut
  {NoStop}%
\bibitem [{\citenamefont {Lee}(2016)}]{lee_anomalous_2016}%
  \BibitemOpen
  \bibfield  {author} {\bibinfo {author} {\bibfnamefont {T.~E.}\ \bibnamefont
  {Lee}},\ }\bibfield  {title} {\bibinfo {title} {Anomalous edge state in a
  non-{Hermitian} lattice},\ }\href
  {https://doi.org/https://doi.org/10.1103/PhysRevLett.116.133903} {\bibfield
  {journal} {\bibinfo  {journal} {Phys. Rev. Lett.}\ }\textbf {\bibinfo
  {volume} {116}},\ \bibinfo {pages} {133903} (\bibinfo {year}
  {2016})}\BibitemShut {NoStop}%
\bibitem [{\citenamefont {Kunst}\ \emph {et~al.}(2018)\citenamefont {Kunst},
  \citenamefont {Edvardsson}, \citenamefont {Budich},\ and\ \citenamefont
  {Bergholtz}}]{kunst_biorthogonal_2018}%
  \BibitemOpen
  \bibfield  {author} {\bibinfo {author} {\bibfnamefont {F.~K.}\ \bibnamefont
  {Kunst}}, \bibinfo {author} {\bibfnamefont {E.}~\bibnamefont {Edvardsson}},
  \bibinfo {author} {\bibfnamefont {J.~C.}\ \bibnamefont {Budich}},\ and\
  \bibinfo {author} {\bibfnamefont {E.~J.}\ \bibnamefont {Bergholtz}},\
  }\bibfield  {title} {\bibinfo {title} {Biorthogonal {Bulk}-{Boundary}
  {Correspondence} in {Non}-{Hermitian} {Systems}},\ }\href
  {https://doi.org/10.1103/PhysRevLett.121.026808} {\bibfield  {journal}
  {\bibinfo  {journal} {Phys. Rev. Lett.}\ }\textbf {\bibinfo {volume} {121}},\
  \bibinfo {pages} {026808} (\bibinfo {year} {2018})}\BibitemShut {NoStop}%
\bibitem [{\citenamefont {Borgnia}\ \emph {et~al.}(2020)\citenamefont
  {Borgnia}, \citenamefont {Kruchkov},\ and\ \citenamefont
  {Slager}}]{Slager2020}%
  \BibitemOpen
  \bibfield  {author} {\bibinfo {author} {\bibfnamefont {D.~S.}\ \bibnamefont
  {Borgnia}}, \bibinfo {author} {\bibfnamefont {A.~J.}\ \bibnamefont
  {Kruchkov}},\ and\ \bibinfo {author} {\bibfnamefont {R.-J.}\ \bibnamefont
  {Slager}},\ }\bibfield  {title} {\bibinfo {title} {Non-hermitian boundary
  modes and topology},\ }\href {https://doi.org/10.1103/PhysRevLett.124.056802}
  {\bibfield  {journal} {\bibinfo  {journal} {Phys. Rev. Lett.}\ }\textbf
  {\bibinfo {volume} {124}},\ \bibinfo {pages} {056802} (\bibinfo {year}
  {2020})}\BibitemShut {NoStop}%
\bibitem [{\citenamefont {Shen}\ \emph {et~al.}(2018)\citenamefont {Shen},
  \citenamefont {Zhen},\ and\ \citenamefont {Fu}}]{shen_topological_2018}%
  \BibitemOpen
  \bibfield  {author} {\bibinfo {author} {\bibfnamefont {H.}~\bibnamefont
  {Shen}}, \bibinfo {author} {\bibfnamefont {B.}~\bibnamefont {Zhen}},\ and\
  \bibinfo {author} {\bibfnamefont {L.}~\bibnamefont {Fu}},\ }\bibfield
  {title} {\bibinfo {title} {Topological band theory for non-hermitian
  hamiltonians},\ }\href {https://doi.org/10.1103/PhysRevLett.120.146402}
  {\bibfield  {journal} {\bibinfo  {journal} {Phys. Rev. Lett.}\ }\textbf
  {\bibinfo {volume} {120}},\ \bibinfo {pages} {146402} (\bibinfo {year}
  {2018})}\BibitemShut {NoStop}%
\bibitem [{\citenamefont {Childs}(2010)}]{Childs2010}%
  \BibitemOpen
  \bibfield  {author} {\bibinfo {author} {\bibfnamefont {A.~M.}\ \bibnamefont
  {Childs}},\ }\bibfield  {title} {\bibinfo {title} {On the relationship
  between continuous- and discrete-time quantum walk},\ }\href
  {https://doi.org/10.1007/s00220-009-0930-1} {\bibfield  {journal} {\bibinfo
  {journal} {Commun. Math. Phys.}\ }\textbf {\bibinfo {volume} {294}},\
  \bibinfo {pages} {581–603} (\bibinfo {year} {2010})}\BibitemShut {NoStop}%
\bibitem [{\citenamefont {Haroche}\ and\ \citenamefont
  {Raimond}(2006)}]{haroche_6exploring_2006}%
  \BibitemOpen
  \bibfield  {author} {\bibinfo {author} {\bibfnamefont {S.}~\bibnamefont
  {Haroche}}\ and\ \bibinfo {author} {\bibfnamefont {J.-M.}\ \bibnamefont
  {Raimond}},\ }\href@noop {} {\emph {\bibinfo {title} {Exploring the quantum:
  atoms, cavities, and photons}}}\ (\bibinfo  {publisher} {Oxford university
  press},\ \bibinfo {year} {2006})\BibitemShut {NoStop}%
\bibitem [{\citenamefont {Xue}\ \emph {et~al.}(2022)\citenamefont {Xue},
  \citenamefont {Hu}, \citenamefont {Song},\ and\ \citenamefont
  {Wang}}]{xue_non-hermitian_2022}%
  \BibitemOpen
  \bibfield  {author} {\bibinfo {author} {\bibfnamefont {W.-T.}\ \bibnamefont
  {Xue}}, \bibinfo {author} {\bibfnamefont {Y.-M.}\ \bibnamefont {Hu}},
  \bibinfo {author} {\bibfnamefont {F.}~\bibnamefont {Song}},\ and\ \bibinfo
  {author} {\bibfnamefont {Z.}~\bibnamefont {Wang}},\ }\bibfield  {title}
  {\bibinfo {title} {Non-{Hermitian} {Edge} {Burst}},\ }\href
  {https://doi.org/10.1103/PhysRevLett.128.120401} {\bibfield  {journal}
  {\bibinfo  {journal} {Phys. Rev. Lett.}\ }\textbf {\bibinfo {volume} {128}},\
  \bibinfo {pages} {120401} (\bibinfo {year} {2022})}\BibitemShut {NoStop}%
\bibitem [{\citenamefont {Cohen-Tannoudji}\ \emph {et~al.}(1998)\citenamefont
  {Cohen-Tannoudji}, \citenamefont {Dupont-Roc},\ and\ \citenamefont
  {Grynberg}}]{cohen_atom_1998}%
  \BibitemOpen
  \bibfield  {author} {\bibinfo {author} {\bibfnamefont {C.}~\bibnamefont
  {Cohen-Tannoudji}}, \bibinfo {author} {\bibfnamefont {J.}~\bibnamefont
  {Dupont-Roc}},\ and\ \bibinfo {author} {\bibfnamefont {G.}~\bibnamefont
  {Grynberg}},\ }\href@noop {} {\emph {\bibinfo {title} {Atom-photon
  interactions: basic processes and applications}}}\ (\bibinfo  {publisher}
  {John Wiley \& Sons},\ \bibinfo {year} {1998})\BibitemShut {NoStop}%
\bibitem [{\citenamefont {Dür}\ \emph {et~al.}(2000)\citenamefont {Dür},
  \citenamefont {Vidal},\ and\ \citenamefont {Cirac}}]{dur_three_2000}%
  \BibitemOpen
  \bibfield  {author} {\bibinfo {author} {\bibfnamefont {W.}~\bibnamefont
  {Dür}}, \bibinfo {author} {\bibfnamefont {G.}~\bibnamefont {Vidal}},\ and\
  \bibinfo {author} {\bibfnamefont {J.~I.}\ \bibnamefont {Cirac}},\ }\bibfield
  {title} {\bibinfo {title} {Three qubits can be entangled in two inequivalent
  ways},\ }\href {https://doi.org/10.1103/PhysRevA.62.062314} {\bibfield
  {journal} {\bibinfo  {journal} {Phys. Rev. A}\ }\textbf {\bibinfo {volume}
  {62}},\ \bibinfo {pages} {062314} (\bibinfo {year} {2000})}\BibitemShut
  {NoStop}%
\bibitem [{\citenamefont {Rudner}\ and\ \citenamefont
  {Levitov}(2009)}]{rudner_topological_2009}%
  \BibitemOpen
  \bibfield  {author} {\bibinfo {author} {\bibfnamefont {M.~S.}\ \bibnamefont
  {Rudner}}\ and\ \bibinfo {author} {\bibfnamefont {L.~S.}\ \bibnamefont
  {Levitov}},\ }\bibfield  {title} {\bibinfo {title} {Topological {Transition}
  in a {Non}-{Hermitian} {Quantum} {Walk}},\ }\href
  {https://doi.org/10.1103/PhysRevLett.102.065703} {\bibfield  {journal}
  {\bibinfo  {journal} {Phys. Rev. Lett.}\ }\textbf {\bibinfo {volume} {102}},\
  \bibinfo {pages} {065703} (\bibinfo {year} {2009})}\BibitemShut {NoStop}%
\bibitem [{\citenamefont {Perczel}\ \emph {et~al.}(2017)\citenamefont
  {Perczel}, \citenamefont {Borregaard}, \citenamefont {Chang}, \citenamefont
  {Pichler}, \citenamefont {Yelin}, \citenamefont {Zoller},\ and\ \citenamefont
  {Lukin}}]{perczel_topological_2017}%
  \BibitemOpen
  \bibfield  {author} {\bibinfo {author} {\bibfnamefont {J.}~\bibnamefont
  {Perczel}}, \bibinfo {author} {\bibfnamefont {J.}~\bibnamefont {Borregaard}},
  \bibinfo {author} {\bibfnamefont {D.~E.}\ \bibnamefont {Chang}}, \bibinfo
  {author} {\bibfnamefont {H.}~\bibnamefont {Pichler}}, \bibinfo {author}
  {\bibfnamefont {S.~F.}\ \bibnamefont {Yelin}}, \bibinfo {author}
  {\bibfnamefont {P.}~\bibnamefont {Zoller}},\ and\ \bibinfo {author}
  {\bibfnamefont {M.~D.}\ \bibnamefont {Lukin}},\ }\bibfield  {title} {\bibinfo
  {title} {Topological quantum optics in two-dimensional atomic arrays},\
  }\href {https://doi.org/10.1103/PhysRevLett.119.023603} {\bibfield  {journal}
  {\bibinfo  {journal} {Phys. Rev. Lett.}\ }\textbf {\bibinfo {volume} {119}},\
  \bibinfo {pages} {023603} (\bibinfo {year} {2017})}\BibitemShut {NoStop}%
\bibitem [{\citenamefont {Okuma}\ \emph {et~al.}(2020)\citenamefont {Okuma},
  \citenamefont {Kawabata}, \citenamefont {Shiozaki},\ and\ \citenamefont
  {Sato}}]{okuma_topological_2020}%
  \BibitemOpen
  \bibfield  {author} {\bibinfo {author} {\bibfnamefont {N.}~\bibnamefont
  {Okuma}}, \bibinfo {author} {\bibfnamefont {K.}~\bibnamefont {Kawabata}},
  \bibinfo {author} {\bibfnamefont {K.}~\bibnamefont {Shiozaki}},\ and\
  \bibinfo {author} {\bibfnamefont {M.}~\bibnamefont {Sato}},\ }\bibfield
  {title} {\bibinfo {title} {Topological {Origin} of {Non}-{Hermitian} {Skin}
  {Effects}},\ }\href {https://doi.org/10.1103/PhysRevLett.124.086801}
  {\bibfield  {journal} {\bibinfo  {journal} {Phys. Rev. Lett.}\ }\textbf
  {\bibinfo {volume} {124}},\ \bibinfo {pages} {086801} (\bibinfo {year}
  {2020})}\BibitemShut {NoStop}%
\bibitem [{\citenamefont {Zhang}\ \emph {et~al.}(2020)\citenamefont {Zhang},
  \citenamefont {Yang},\ and\ \citenamefont
  {Fang}}]{zhang_correspondence_2020}%
  \BibitemOpen
  \bibfield  {author} {\bibinfo {author} {\bibfnamefont {K.}~\bibnamefont
  {Zhang}}, \bibinfo {author} {\bibfnamefont {Z.}~\bibnamefont {Yang}},\ and\
  \bibinfo {author} {\bibfnamefont {C.}~\bibnamefont {Fang}},\ }\bibfield
  {title} {\bibinfo {title} {Correspondence between {Winding} {Numbers} and
  {Skin} {Modes} in {Non}-{Hermitian} {Systems}},\ }\href
  {https://doi.org/10.1103/PhysRevLett.125.126402} {\bibfield  {journal}
  {\bibinfo  {journal} {Phys. Rev. Lett.}\ }\textbf {\bibinfo {volume} {125}},\
  \bibinfo {pages} {126402} (\bibinfo {year} {2020})}\BibitemShut {NoStop}%
\bibitem [{\citenamefont {Esaki}\ \emph {et~al.}(2011)\citenamefont {Esaki},
  \citenamefont {Sato}, \citenamefont {Hasebe},\ and\ \citenamefont
  {Kohmoto}}]{esaki_edge_2011}%
  \BibitemOpen
  \bibfield  {author} {\bibinfo {author} {\bibfnamefont {K.}~\bibnamefont
  {Esaki}}, \bibinfo {author} {\bibfnamefont {M.}~\bibnamefont {Sato}},
  \bibinfo {author} {\bibfnamefont {K.}~\bibnamefont {Hasebe}},\ and\ \bibinfo
  {author} {\bibfnamefont {M.}~\bibnamefont {Kohmoto}},\ }\bibfield  {title}
  {\bibinfo {title} {Edge states and topological phases in non-{Hermitian}
  systems},\ }\href
  {https://doi.org/https://doi.org/10.1103/PhysRevB.84.205128} {\bibfield
  {journal} {\bibinfo  {journal} {Phys. Rev. B}\ }\textbf {\bibinfo {volume}
  {84}},\ \bibinfo {pages} {205128} (\bibinfo {year} {2011})}\BibitemShut
  {NoStop}%
\bibitem [{\citenamefont {Sato}\ \emph {et~al.}(2012)\citenamefont {Sato},
  \citenamefont {Hasebe}, \citenamefont {Esaki},\ and\ \citenamefont
  {Kohmoto}}]{sato_time-reversal_2012}%
  \BibitemOpen
  \bibfield  {author} {\bibinfo {author} {\bibfnamefont {M.}~\bibnamefont
  {Sato}}, \bibinfo {author} {\bibfnamefont {K.}~\bibnamefont {Hasebe}},
  \bibinfo {author} {\bibfnamefont {K.}~\bibnamefont {Esaki}},\ and\ \bibinfo
  {author} {\bibfnamefont {M.}~\bibnamefont {Kohmoto}},\ }\bibfield  {title}
  {\bibinfo {title} {Time-{Reversal} {Symmetry} in {Non}-{Hermitian}
  {Systems}},\ }\href {https://doi.org/10.1143/PTP.127.937} {\bibfield
  {journal} {\bibinfo  {journal} {Prog. Theor. Phys.}\ }\textbf {\bibinfo
  {volume} {127}},\ \bibinfo {pages} {937} (\bibinfo {year}
  {2012})}\BibitemShut {NoStop}%
\bibitem [{\citenamefont {Kawabata}\ \emph
  {et~al.}(2019{\natexlab{b}})\citenamefont {Kawabata}, \citenamefont
  {Higashikawa}, \citenamefont {Gong}, \citenamefont {Ashida},\ and\
  \citenamefont {Ueda}}]{kawabata_topological_2019}%
  \BibitemOpen
  \bibfield  {author} {\bibinfo {author} {\bibfnamefont {K.}~\bibnamefont
  {Kawabata}}, \bibinfo {author} {\bibfnamefont {S.}~\bibnamefont
  {Higashikawa}}, \bibinfo {author} {\bibfnamefont {Z.}~\bibnamefont {Gong}},
  \bibinfo {author} {\bibfnamefont {Y.}~\bibnamefont {Ashida}},\ and\ \bibinfo
  {author} {\bibfnamefont {M.}~\bibnamefont {Ueda}},\ }\bibfield  {title}
  {\bibinfo {title} {Topological unification of time-reversal and particle-hole
  symmetries in non-{Hermitian} physics},\ }\href
  {https://doi.org/https://doi.org/10.1038/s41467-018-08254-y} {\bibfield
  {journal} {\bibinfo  {journal} {Nat. Commun.}\ }\textbf {\bibinfo {volume}
  {10}},\ \bibinfo {pages} {297} (\bibinfo {year}
  {2019}{\natexlab{b}})}\BibitemShut {NoStop}%
\bibitem [{\citenamefont {Chiu}\ \emph {et~al.}(2016)\citenamefont {Chiu},
  \citenamefont {Teo}, \citenamefont {Schnyder},\ and\ \citenamefont
  {Ryu}}]{chiu_classification_2016}%
  \BibitemOpen
  \bibfield  {author} {\bibinfo {author} {\bibfnamefont {C.-K.}\ \bibnamefont
  {Chiu}}, \bibinfo {author} {\bibfnamefont {J.~C.~Y.}\ \bibnamefont {Teo}},
  \bibinfo {author} {\bibfnamefont {A.~P.}\ \bibnamefont {Schnyder}},\ and\
  \bibinfo {author} {\bibfnamefont {S.}~\bibnamefont {Ryu}},\ }\bibfield
  {title} {\bibinfo {title} {Classification of topological quantum matter with
  symmetries},\ }\href
  {https://doi.org/https://doi.org/10.1103/RevModPhys.88.035005} {\bibfield
  {journal} {\bibinfo  {journal} {Rev. Mod. Phys.}\ }\textbf {\bibinfo {volume}
  {88}},\ \bibinfo {pages} {035005} (\bibinfo {year} {2016})}\BibitemShut
  {NoStop}%
\bibitem [{\citenamefont {Haga}\ \emph {et~al.}(2021)\citenamefont {Haga},
  \citenamefont {Nakagawa}, \citenamefont {Hamazaki},\ and\ \citenamefont
  {Ueda}}]{Haga2021}%
  \BibitemOpen
  \bibfield  {author} {\bibinfo {author} {\bibfnamefont {T.}~\bibnamefont
  {Haga}}, \bibinfo {author} {\bibfnamefont {M.}~\bibnamefont {Nakagawa}},
  \bibinfo {author} {\bibfnamefont {R.}~\bibnamefont {Hamazaki}},\ and\
  \bibinfo {author} {\bibfnamefont {M.}~\bibnamefont {Ueda}},\ }\bibfield
  {title} {\bibinfo {title} {Liouvillian skin effect: Slowing down of
  relaxation processes without gap closing},\ }\href
  {https://doi.org/10.1103/PhysRevLett.127.070402} {\bibfield  {journal}
  {\bibinfo  {journal} {Phys. Rev. Lett.}\ }\textbf {\bibinfo {volume} {127}},\
  \bibinfo {pages} {070402} (\bibinfo {year} {2021})}\BibitemShut {NoStop}%
\end{thebibliography}%

\appendix

\section{Topological invariant for short-range chiral-symmetric systems}

In the main text, we have established that the long-range hopping two-band model $\tilde{H}'_{\text{TB}}(k)$ has the SSH-like form %Eq 
(\ref{eq:twoband_SSH}), with a quasi-momentum $kd\in[-\pi/2,\pi/2]$ and a discontinuous off-diagonal matrix element $g(k)$. Here, we consider the topological invariant of its short-range analogy $H_{\text{SR}}$, where the quasi-momentum is defined as $kd\in[-\pi,\pi]$ and the off-diagonal element $g(k)=g$ being a real constant. 

Point-gapped non-Hermitian systems encounter novel features without Hermitian counterparts, such as the skin effect \cite{gong_topological_2018, okuma_topological_2020, zhang_correspondence_2020}. However, a line-gapped system can always %the line-gap spectrum can 
be mapped onto a %\zpg{gapped [A metal also has a Fermi level but is gapless]} 
Hermitian system, as the case here~\cite{esaki_edge_2011, kawabata_symmetry_2019, ashida_non-hermitian_2020}. To perform the Hermitization of the two-band model, we first note that $i\sigma_{x}, i\sigma_{y}$ and $\sigma_{z}$ forms a $SU(1,1)$ algebra~\cite{sato_time-reversal_2012}. Consequently, $H_{\text{TB}}(k)$ exhibits the pseudo-Hermiticity: 
\begin{equation*}
\eta H_{\text{SR}}(k) \eta = H_{\text{SR}}(k)^{\dagger}, 
\quad \eta = \sigma_{z}. 
\end{equation*}
We multiply $H_{\text{SR}}(k)$ %$\tilde{H}_{\text{TB}}'(k)$ 
by $i$ \cite{kawabata_topological_2019}
\begin{equation}
iH_{\text{SR}}(k)=ih\sigma_z + g\cos(kd)\sigma_x + g\sin(kd)\sigma_y
%i\tilde{H}_{\text{TB}}'(k) =  ih \sigma_{z} + g(k)\cos(kd)\sigma_{x } + g(k)\sin(kd)\sigma_{y},
\label{eq:two_band_linegapped}
\end{equation}
to turn the imaginary line gap into a real gap. The pseudo-Hermiticity transforms into pseudo-anti-Hermiticity, which may also be called chiral symmetry \cite{kawabata_symmetry_2019}:
\begin{equation*}
\eta H_{\text{SR}}(k) \eta = -H_{\text{SR}}(k)^{\dagger}
%\eta H \eta = -H^{\dagger}, 
\quad \eta = \sigma_{z}. 
\end{equation*}

The winding number of $iH_{\text{SR}}(k)$ can be defined by using the following projection operators:
\begin{equation*}
P_{1}(k) =  |R_{+}(k)\rangle \langle L_{+}(k)|, \quad P_{2}(k) =  |L_{-}(k)\rangle \langle R_{-}(k)|,
\end{equation*}
where $|R_{\pm}(k)\rangle$ ($|L_{\pm}(k)\rangle$) are the right (left) eigenvectors of $H_{\text{SR}}(k)$, 
to construct a Hermitian operator
\begin{equation}
\begin{split}
Q(k) &= 1 - \left( P_{1}(k) + P_{2}(k) \right) = \frac{1}{\sqrt{g^{2} - h^{2}}} \begin{pmatrix}
0 & ge^{-ikd} \\
ge^{ikd} & 0\\
\end{pmatrix}\\
&=\begin{pmatrix}
0 & q(k) \\
q^{*}(k) & 0\\
\end{pmatrix},
\end{split}
\label{eq:doubled}
\end{equation}
which is Hermitian and chiral symmetric: 
\begin{equation*}
\Gamma Q(k) \Gamma = -Q(k), \quad \Gamma = \sigma_{z}.
\end{equation*}

Technically speaking, $Q(k)$ belongs to class BDI
of the Altland-Zirnbauer classification~\cite{chiu_classification_2016} and is characterized by the
winding number $\int_{\rm B.Z.}\frac{dk}{ 2\pi %(B. Z.) 
i} q(k)^{-1}\partial_{k}q(k)  \in \mathbb{Z}$ for a continuous $q(k)$. Therefore, the topological invariant for the short-range non-Hermitian SSH model can be defined. 

In the case of our long-range Hamiltonian $H'_{\rm TB}(k)$, we can identify the corresponding $q(k)$ following a similar procedure. However, the discontinuity of $q(k)$ at $k=\pm \pi/(2d)$ prohibits us from concluding a quantized winding number.

\section{Normalization property of $F(x,t)$}
We show the generalized spatially resolved escape distribution $F(x,t)$ is normalized for $t=0$. For a normalized initial state $\langle\psi_0|\psi_0\rangle = 1$ and $H_{\text{TB}}$ with a complex spectrum $\{E_{n}\}$ that lies only on the lower half of the complex energy plane, summing over space gives
\begin{equation*}
\begin{split}
\sum_{x} F(x,0) &= -\sum_{x}\int_0^\infty dt'\langle\psi(t')|\{|x\rangle\langle x|,{\rm Im}H_{\text{TB}} \}|\psi(t')\rangle \\
&= -2 \int_0^\infty dt'\langle\psi(t')|{\rm Im}H_{\text{TB}} |\psi(t')\rangle \\
&= - \int_0^\infty dt' \bra{\psi(t')}iH_{\text{TB}}^\dag-iH_{\text{TB}}\ket{\psi(t')}  \\
&= - \int_0^\infty dt' \frac{d}{dt'}\bra{\psi(t')}\ket{\psi(t')} \\
& =\langle\psi_0|\psi_0\rangle - \langle\psi_\infty|\psi_\infty\rangle =1, 
\end{split}
\end{equation*}
where $ |\psi_\infty \rangle = \lim_{t\to\infty} e^{-iH_{\rm TB}t}\ket{\psi_0} =0$ regardless the diagonalizability of $H_{\rm TB}$. To see this, we consider following spectral decomposition of $ H_{\rm TB}$ in the Jordan normal form \cite{ashida_non-hermitian_2020}
\begin{equation*}
\begin{split}
\| e^{-iH_{\rm TB}t} \| &= \left \Vert \sum_{n} e^{-i E_{n} t} \left( P_n + \sum_{i=1}^{m_{n}^{g}} \sum_{j=1}^{n_i -1} \frac{t^{j}}{j!} N_{n_i}^{j} \right) \right \Vert  \\
&\leq  \sum_{n} e^{\Im(E_{n}) t} \left \Vert \left( P_n + \sum_{i=1}^{m_{n}^{g}} \sum_{j=1}^{n_i -1} \frac{t^{j}}{j!} N_{n_i}^{j} \right) \right \Vert,  %\\
%&= 0,
\end{split}
\end{equation*}
where for each Jordan block with eigenvalue $E_{n}$, $P_n$ is the set of orthogonal and complete projectors, $m_{n}^{g}$ is the geometric multiplicities, and $N_{n_i}$ is a nilpotent off-diagonal block with size $n_i$. Because $\{\Im(E)\}$ lies below the real axis, the exponential term decreases faster than the polynomial growth, and thus the above equation vanishes in the limit of $t\to\infty$. We mention that it is necessary to assume a finite system and take this infinite time limit first. Otherwise, if the sizes of some Jordan blocks grow with the system size, the polynomial contribution may dramatically enlarge the relaxation time, possibly to infinity in the thermodynamic limit. This occurs in some systems exhibiting the non-Hermitian skin effect \cite{Haga2021}.

\end{document}